\RequirePackage{fix-cm}
\documentclass[twocolumn,epjc3]{svjour3}
\smartqed  % flush right qed marks, e.g. at end of proof
\RequirePackage{graphicx}
\usepackage{color}
\usepackage{amssymb}
\usepackage{amsmath}
\usepackage{xspace}
\journalname{Eur. Phys. J. C}
\begin{document}

\title{Particle Physics and Cosmology \\
{\it \small Lecture notes from the 100$^{\mathrm{th}}$ Les Houches Summer School on Post-Planck Cosmology, July 8$^{th}$ - Aug 2$^{nd}$ 2013}}


\author{P. Pralavorio\thanksref{e2,addr2}
}

\thankstext{e2}{e-mail: pascal.pralavorio@cern.ch}


\institute{CPPM, Univ. Aix-Marseille and CNRS/IN2P3, 163 avenue de Luminy, case 902, 13288 Marseille cedex 09 (France) \label{addr2}
}

\date{}

\maketitle

\begin{abstract}
Today, both particle physics and cosmology are described by few parameter Standard Models, i.e. it is possible to deduce consequence of particle physics in 
cosmology and vice versa. The former is examined in this lecture, in light of the recent systematic exploration of the electroweak scale by the LHC 
experiments. The two main results of the first phase of the LHC, the discovery of a Higgs-like particle and the absence so far of new particles predicted by 
``natural" theories beyond the Standard Model (supersymmetry, extra-dimension and composite Higgs) are put in a historical context 
to enlighten their importance and then presented extensively. To be complete, a short review from the neutrino physics, which can not be probed at LHC, is also given. 
The ability of all these results to resolve the 3 fundamental questions of cosmology about the nature of dark energy and dark matter as well as the 
origin of matter-antimatter asymmetry is discussed in each case.
\end{abstract}

%&&&&&&&&&&&&&&&&&&&&&&&&&&&&&&&&&&&&&&&&&&&&&&&&&&&&&&&&&&&&&&&&&&&&&&&&&&&&&&&&&&&&&&&&&&&&&&&&&&&&&&&&&&&&&&&&&&&&&&&&&&&&&&&&&&&&
%&&&&&&&&&&&&&&&&&&&&&&&&&&&&&&&&&
%&&&&&&&&&&&&&&&&&&&&&&&&&&&&&&&&&
%&&&&&&&&&&&&&&&&&&&&&&&&&&&&&&&&&
\tableofcontents
\section{Introduction}
\label{Part:Intro}
%&&&&&&&&&&&&&&&&&&&&&&&&&&&&&&&&&

\begin{figure}[htbp]
\begin{center}    
\includegraphics[height=7cm]{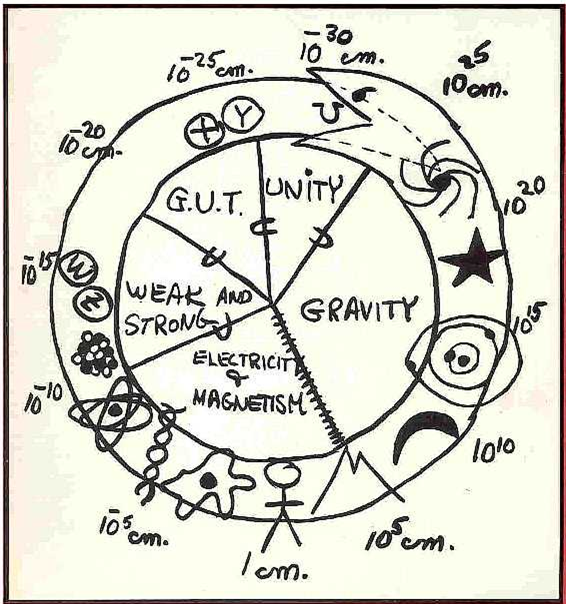}
\end{center}
\caption{S. Glashow serpent swallowing its tail, illustrating the interconnection between particle physics and cosmology~\protect\cite{GLASHOW_GEORGI}.}
\label{fig:Serpent}
\end{figure}

The description of particle physics interactions and gravity in a common framework is still an ongoing effort but 
it was mostly speculative or purely philosophical at the beginning of the XX$^{\mathrm{th}}$ century. The first attempt in this direction was to introduce 
a cosmological constant $\Lambda$ in the theory of gravitation. However the discovery of the universe expansion in the 20's forced 
Einstein to abandon this constant. In the 30's, pioneering works from Lema\^{\i}tre on primeval atom and identification between 
$\Lambda$ and the vacuum energy were poorly received by the community, and particle physics developed apart from
cosmology for more than 30 years. The subject reappeared in a seminal paper entitled ``The Universe as a Hot Laboratory 
for the Nuclear and Particle Physicist" by Zel'dovich in the late 1960s. At the beginning of the 1980s, the construction of Grand 
Unified Theories (GUT) at very high energy scale O($10^{15-16}$) GeV connected finally particle physics and cosmology. 
In simple words, ``(...) the universe is the only machine we have that can test 
these GUT ideas. It's the world's biggest particle accelerator (...). But it's hard to use because all the experiments happened only once, (...) a long time 
ago''~\cite{GLASHOW_GEORGI}. The connection between particle physics and cosmology is nicely illustrated by the cosmic serpent shown in 
Fig.~\ref{fig:Serpent}.

Today, benefiting from the huge technological developments during the last century, gigantic 
underground accelerators and spacecrafts probe everyday these two extreme realizations of physics. 
So what is more precisely the situation of particle physics and cosmology? At the end 
of the last century both fields were mature enough to give birth to ``Standard Models". Both only need few free parameters to explain the 
huge wealth of data even if most of these parameters could not be explained yet in terms of a fundamental theory. 
But what is new is that these models enable to deduce consequence of particle physics in 
cosmology and vice verse, one of the most outstanding result of modern science. As confirmed by the latest Planck results, cosmology provides three 
fundamental questions on today's universe:
\begin{enumerate}
\item What is dark energy? 
\item What is dark matter?
\item Why is there matter and not antimatter (baryogenesis)? 
\end{enumerate}
May be part (or all) of these questions can be resolved by particle physics. Section~\ref{Part:HiggsSM} discusses, in light of the recent collider results, 
how particle physics could contribute to the solution of the first question. Section~\ref{sec:chap2} presents the status of the collider and 
neutrino experiment searches for new particles beyond the Standard Model that could be able to solve the two last questions. 
Section~\ref{Part:Conclusion} concludes by giving a tentative answer to these three questions using particle physics.

Particle physics results discussed in this lecture are a digest of the Large Hadron Collider (LHC) experiment publications by July 2013. At LHC, proton-proton collisions 
were recorded between spring 2010 and autumn 2012. More than 500 `LHC' papers were produced by July 2013 (comparable to the production of the last 10 years in 
Particle physics before LHC start). The quantity of data recorded by general purpose experiments ATLAS and CMS are proportional to the integrated luminosity $L$,
expressed in fb$^{-1}$, as the number $N$ of events expected for a given process of cross-section $\sigma$ is $N=L\sigma$. Cross-sections are 
expressed in pb ($10^{-36}~\mathrm{cm}^2$) or fb ($10^{-39}~\mathrm{cm}^2$). LHC accumulated 5~fb$^{-1}$ and 20~fb$^{-1}$ of data at a center of mass energy 
$\sqrt{s}=7$ TeV and 8 TeV, respectively. This two data sets are referred to as `LHC Run I'. 
In this lecture, energy is expressed in MeV ($10^{-3}$ GeV), GeV or TeV ($10^3$ GeV). It corresponds to temperature and to the inverse of a 
distance (m$^{-1}$) and time (s$^{-1}$) when using natural units $\hbar = c = k_B = 1$. Useful physical or astrophysical constants are the Planck mass, 
$M_{Pl}=\sqrt{\hbar c/G_N} \simeq 1.2 \times 10^{19}$ GeV, where $G_N$ is the Newtonian gravitational constant, the Fermi coupling constant, 
$G_F/(\hbar c)^3 \simeq 1.2 \times 10^{-5}~\mathrm{GeV}^{-2}$, which characterizes the weak interaction and the fine-structure 
constant, $\alpha=e^2/(4\pi \varepsilon_0 \hbar c) \simeq 7.3 \times 10^{-3}$, which characterizes the electromagnetic (EM) interaction.

%&&&&&&&&&&&&&&&&&&&&&&&&&&&&&&&&&&&&&&&&&&&&&&&&&&&&&&&&&&&&&&&&&&&&&&&&&&&&&&&&&&&&&&&&&&&&&&&&&&&&&&&&&&&&&&&&&&&&&&&&&&&&&&&&&&&&
%&&&&&&&&&&&&&&&&&&&&&&&&&&&&&&&&&
%&&&&&&&&&&&&&&&&&&&&&&&&&&&&&&&&&
%&&&&&&&&&&&&&&&&&&&&&&&&&&&&&&&&&
\section{Standard Model of Particle Physics and Cosmology}
\label{Part:HiggsSM}
%&&&&&&&&&&&&&&&&&&&&&&&&&&&&&&&&&

\begin{figure*}[htbp]
\begin{center}    
\includegraphics[height=10cm]{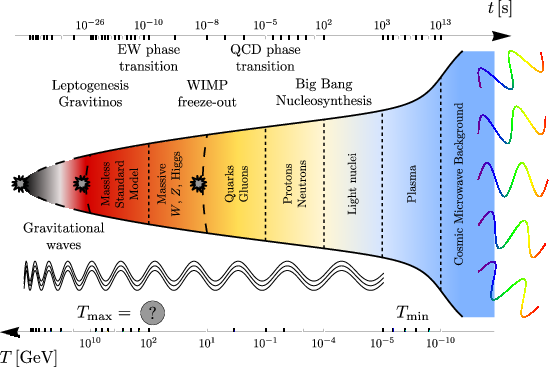}
\end{center}
\caption{Chronology of the hot thermal phase of the early universe. The top scale shows the time after the Big Bang and the bottom scale the corresponding 
post-inflationary reheating temperature~\protect\cite{CosmoPartPhysics_Fig}.}
\label{fig:ParticleCosmo1}
\end{figure*}

Our current best guess of the early universe and its connection with particle physics is presented in Fig.~\ref{fig:ParticleCosmo1}. It is very 
interesting to note that only the last epochs of the early universe, corresponding to 
$10^{-2} < t~(\mathrm{s}) < 10^{13}$ ($10^{-10} < T_R~(\mathrm{GeV}) < 10^{-2}$), have been probed experimentally by Big Bang nucleosynthesis (BBN) and 
cosmic microwave background (CMB). This tells us that the post-inflationary reheating temperature $T_R$ of the universe have reached at least 
10 MeV~\cite{CosmoPartPhysics}. It is of course possible to access indirectly hotter epochs, but their direct 
exploration will require the very challenging detection of the cosmic neutrino background (CvB) and/or gravitational waves. 

With the advent of particle physics colliders at the end of the 50's, it is however possible to produce experimentally energies much higher than 10 MeV. 
The most powerful one today is the LHC with $\sqrt{s}=8$ TeV, 6 order of magnitude beyond what BBN can probe. Therefore developments in particle physics will 
directly benefit to cosmology -- even if it is complicated by the irrelevance of gravity in particle physics experiments. It goes without saying that 
much higher energy collisions happen everyday in the universe (even on earth via cosmic ray interaction with the atmosphere), but particle physics colliders 
provide a unique setting for systematics studies.
 
After a brief historical review of particle physics and earlier connections with cosmology in Section~\ref{sec:PP_Cosmo}, the basics of the 
Standard Model (SM) of Particle Physics as well as LHC collider and experiments are reminded in Section~\ref{sec:SM} and~\ref{sec:LHC}, respectively. An overview of the 
Higgs-like particle discovery and its consequences for the early universe follows in Section~\ref{sec:Higgs_Cosmo}. Finally an explanation of 
the current SM limitations and its possible extensions are given in Section~\ref{sec:Higgs_NP}.

\begin{figure}[htbp]
\begin{center}    
\includegraphics[height=5.5cm]{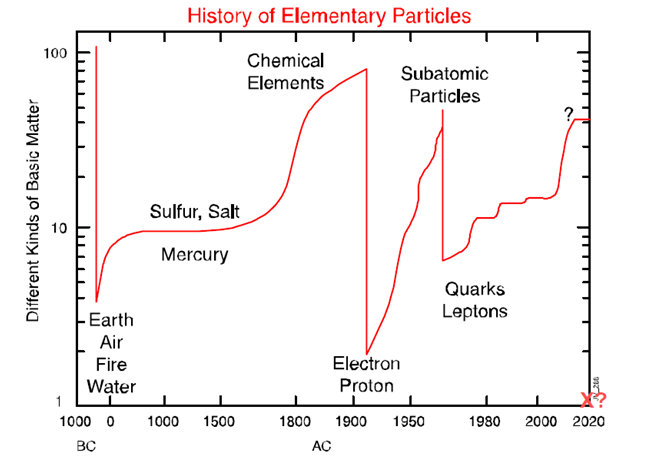}
\end{center}
\caption{Variation of the number of elementary particles as a function of time.}
\label{fig:ParticleCosmo2}
\end{figure}

%&&&&&&&&&&&&&&&&&&&&&&&&&&&&&&&&&
\subsection{History of Particle Physics and Cosmology}
\label{sec:PP_Cosmo}
%&&&&&&&&&&&&&&&&&&&&&&&&&&&&&&&&&

Figure~\ref{fig:ParticleCosmo2} shows the evolution of the number of elementary particles as a function of time. Three dramatic changes can be 
seen: the first one begins at the time of the ancient Greeks, who classified basic matter in four categories (Earth, Air, Fire and Water) and 
culminates when Mendeleev invented the periodic table of elements. A second era opens at the beginning of XX$^{\mathrm{th}}$ century when 
Niels Bohr described the hydrogen atom with two recently discovered particles: the proton~\footnote{In fact the atomic nucleus at the time N. Bohr wrote his paper.} 
and the electron~\cite{BOHR1,BOHR2}. He computed the radius of the atom as: 
\begin{equation}
a_0 = \frac{4 \pi \varepsilon_0 \hbar^2}{m_{\mathrm{e}} e^2} = \frac{\hbar}{m_{\mathrm{e}}\,c\,\alpha} \sim \mathrm{O}(10^{-10})~\mathrm{m}
\label{eq:BohrRadius}
\end{equation}
At this atomic scale, lots of new phenomena were discovered and a new description of microscopic nature, quantum mechanics, emerges in the 20's. Later, 
thanks to the development of detection technology, lots of new particles were discovered first by studying cosmic rays and then as 
decay products of colliding particles. Late in the 60's it was finally possible to make sense of all these new particles with 
quantum field and gauge theories. More specifically, the unification of electromagnetism and 
weak interaction~\cite{SM1,SM2,SM3} including the Higgs mechanism~\cite{HIGGS1,HIGGS2,HIGGS3,HIGGS4,HIGGS5,HIGGS6} allowed a coherent 
description of leptons~\cite{RenormSM}. Similar evolution took place in the hadronic sector with the quark model~\cite{QUARKS} and the description of 
the strong force~\cite{QCD1,QCD2,QCD3}. Hence the third era of elementary particles. 
Spectacularly all particles discovered until then could be explained in terms of these new elementary particles. The Model was completed 
by the prediction of 3 copies of a 2-fold fermion family~\cite{3Family} and the discovery of weak neutral currents~\cite{GarGamelle1,Gargamelle2}, 
a direct consequence of the EM and weak unification. J. Ilioupolos announced officially the birth of the Standard Model, later called SM, with 19 parameters 
at the ICHEP conference in London in July 1974. 
Since then, the SM has been beautifully confirmed by all experiments. All new particles discovered fit perfectly in the model 
framework~\footnote{It is noticeable that all fermion discoveries were made in the US whereas the boson discoveries were made in Europe !}: the leptons 
tau~\cite{TAUDISC} and neutrino-tau~\cite{NTAUDISC}, the quarks charm~\cite{CHARMDISC1,CHARMDISC2}, bottom~\cite{BOTTOMDISC} and 
top~\cite{TOPDISC1,TOPDISC2}, as well as the gluon~\cite{GLUONDISC1,GLUONDISC2,GLUONDISC3,GLUONDISC4}, $W$~\cite{WDISC1,WDISC2}, 
$Z$~\cite{ZDISC1,ZDISC2} and recently the Higgs boson~\cite{HIGGSDISC1,HIGGSDISC2}. 

Cosmology has followed similar historical evolution with abrupt changes: the Geocentric model (Aristotle) was turned into the Heliocentric model with 
elliptic orbits (Copernicus, Kepler). Later, the static uniform universe of Newton, Descartes and Kant, based on the classical gravity 
was turned into a static uniform universe based on General Relativity by Einstein in 1916. These static models were superseded by the dynamical hot Big Bang 
Model of Friedmann-Lemaitre in the 20's, complemented by the theory of inflation in the early 80's which solved the horizon, flatness and magnetic-monopole problems. 
Since 1998 and the concordant results obtained by SuperNovae, CMB and BAO experiments, it is considered as the Standard Model of Cosmology ($\Lambda$CDM) with 6 free 
parameters. 

There are numerous examples of cross-feeding between cosmology and particle physics. The most striking one dates back to 1900 when 
Max Planck postulated the quantization of the energy ($E=h\nu$) to describe correctly the black body radiation. For particle physics, this was a first 
seed in the quantum mechanics garden that will flourish during the XX$^{\mathrm{th}}$ century. For cosmology, this provided a very strong argument in favor of the 
Big Bang theory in the 70's when the CMB temperature spectrum followed exactly the Planck's black body description with $T \sim 2.7$ K. More recently 
nucleosynthesis and the constraints on the mass and number of neutrinos are also good examples of this fruitful connection.

%&&&&&&&&&&&&&&&&&&&&&&&&&&&&&&&&&
\subsection{Standard Model of Particle Physics}
\label{sec:SM}
%&&&&&&&&&&&&&&&&&&&&&&&&&&&&&&&&&

The main features of the current version of the SM of Particle Physics are now given. A more thorough review can be found for example 
in~\cite{SM_Altarelli}. The Standard Model is a non abelian gauge field theory based on the symmetry group 
$SU(3)_C \otimes SU(2)_L \otimes U(1)_Y$. $SU(3)_C$ denotes the color ($C$) group of Quantum Chromo Dynamics (QCD). $SU(2)_L \otimes U(1)_Y$ describes 
the electroweak (EW) interactions where the weak hypercharge $Y$ is the U(1) generator and can be linked to the electric charge ($Q$) and the Weak 
Isospin ($T_3$) by the formula $Y=2(Q-T_3)$. In total, the SM counts 58 objects, 118 degrees of freedom and 28 free parameters, that will be detailed in this section. 

\subsubsection{SM Fields}

\begin{figure}[htbp]
\begin{center}    
\includegraphics[height=3.5cm]{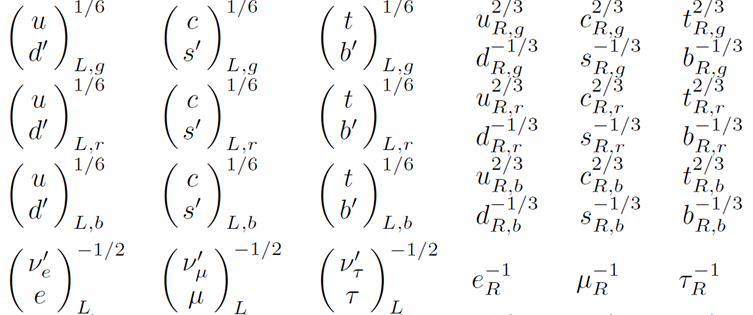}
\end{center}
\caption{SM fermion fields with their associated charges ($Y$ top right, $C=g,r,b$ bottom right of each doublet or singlet) and chirality 
(bottom right of each doublet or singlet)~\protect\cite{Murayama_SUSY}.}
\label{fig:StandardModel1}
\end{figure}

\begin{figure}[htbp]
\begin{center}    
\includegraphics[height=2.5cm]{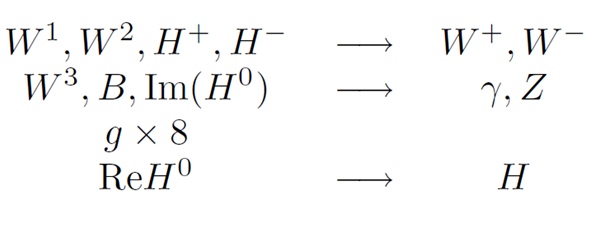}
\end{center}
\caption{SM boson fields before and after EW symmetry breaking~\protect\cite{Murayama_SUSY}.}
\label{fig:StandardModel2}
\end{figure}

The Standard Model includes 45 massive fermion fields arranged in left-handed SU(2) doublets and right-handed SU(2) singlets as shown in 
Fig.~\ref{fig:StandardModel1}. Since parity is maximally violated by the weak interaction~\cite{EW_PViolation}, there is no right-handed neutrinos 
and only left-handed fermions (and right handed anti-fermions) are sensitive to the weak interaction. The primes on down-type 
quarks and neutrinos correspond to gauge eigenstates. They are linked to the mass eigenstates by two 3x3 mixing matrices called Cabibbo-Kobayashi-Maskawa 
(CKM) and Pontecorvo-Maki-Nakagawa-Sakata (PMNS) respectively. 
There are also 12 gauge boson fields corresponding to the different force carriers of the interactions. Nine are massless, the photon and the 8 colored
gluons for electromagnetic and strong interactions respectively. Because it is a short-range force, the gauge bosons of the weak interaction need to be 
very massive, of the order of the weak scale. However this is not possible since $SU(2)_L \otimes U(1)_Y$ is conserved. A solution is provided by the Higgs mechanism.
A complex scalar SU(2) doublet $\phi \equiv \left(\begin{array}{c} \phi^+ \\ \phi^0  \end{array} \right)$ is introduced with a tree-level potential:
\begin{equation}
V(\phi^{\dag}\phi) = -\frac{m_H^2}{2} \phi^{\dag}\phi + \lambda (\phi^{\dag}\phi)^2
\label{eq:HiggsPotential}
\end{equation}
With $-m_H^2 < 0$ and $\lambda > 0$, $V$ has a "Mexican hat" shape with an infinity of non trivial vacua. The vacuum expectation 
value (vev) of $\phi$ can be expressed as:
\begin{equation}
v=\sqrt{m_H^2/(2\lambda)}=(\sqrt{2}G_F)^{-1/2} \simeq 246~\mathrm{GeV}
\label{eq:HiggsVev}
\end{equation}
and $\phi$ can be developped as $\left(\begin{array}{c} H^+ \\ \frac{1}{\sqrt{2}}(v+\mathrm{Re}(H^0)+i\mathrm{Im}(H^0) \end{array} \right)$. 
The Higgs potential is completely determined once $m_H$ is known. By choosing a particular ground state, the gauge symmetry 
$SU(2)_L \otimes U(1)_Y$ gets spontaneously broken in U(1)$_{Q}$. The massless Goldstone bosons ($H^+$, $H^-$ and Im($H^0$)) generated by this symmetry breaking 
mix with the massless gauge bosons $W^{1,2,3}$, as illustrated in Fig.~\ref{fig:StandardModel2}. Because of the Higgs vev, the Higgs mechanism generates three massive bosons 
($W^+$, $W^-$, $Z^0$) for the weak interaction, one massless boson for the EM interaction ($\gamma$) and one massive scalar particle ($H$, the Higgs boson).

\subsubsection{SM Parameters}

Let's now enumerate the free parameters of the model. The interaction between scalar and Dirac fields is described by Yukawa-type Lagrangian 
$\mathcal{L}=-\lambda_Y\bar{\psi} \phi \psi$. It can be rewritten as $-m/v(\bar{\psi_R}\psi_L +\bar{\psi_L}\psi_R)^*$, where $\lambda_Y$ is a real coupling constant, 
$m$ is the Dirac mass and $\psi_L$ ($\psi_R$) are the left-handed (right-handed) spinors. In the fermion sector, Yukawa-type couplings 
give 12 free parameters corresponding to the fermions masses. CKM and PMNS matrices 
can be parametrized by 3 angles and 1 CP violating phase each~\cite{Wolfenstein}. 
Assuming that the neutrino is of Majorana type, 2 other parameters are necessary for the PMNS matrix, 
giving a total of 10 parameters for the fermion mixing matrices. In the boson sector, the electroweak part needs 4 parameters ($\alpha$, $m_Z$, $v$, $m_H$ for example) 
and the strong sector 2 parameters ($\alpha_S$ coupling constant and a strong CP phase, $\Theta_{\mathrm{QCD}}$, very small but not null). 

Overall most of the 28 fundamental SM parameters~\footnote{Originally, the neutrinos were assumed massless and the PMNS matrix diagonal, hence the 19 parameters 
mentioned in Section~\ref{sec:PP_Cosmo}.} are a consequence of the presence of the Higgs field. Before 2012, all these parameters have been measured by experiments, 
except 7 ($m_H$, $\Theta_{\mathrm{QCD}}$, the two Majorana phases of the PMNS matrix and the 3 neutrino masses) which are only constrained. However these constraints 
are generally weak. Therefore final values could have dramatic consequences on cosmology: $m_H$ will be discussed extensively in the following, 
while $\Theta_{\mathrm{QCD}}$ and neutrino parameters are discussed in Section~\ref{sec:DM} and~\ref{sec:neutr}, respectively.

\subsubsection{The Electroweak fit}
\label{sec:EWFit}

\begin{figure}[htbp]
\begin{center}
\includegraphics[height=6cm]{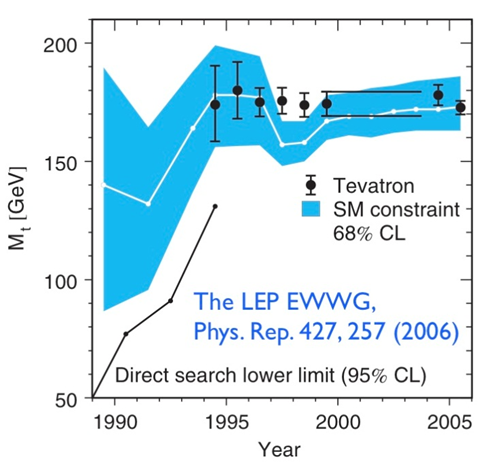}
\end{center}
\caption{Predictions from the EW Fit compared to experimental results for top mass~\protect\cite{EWFit}.}
\label{fig:EWFit1}
\end{figure}

\begin{figure}[htbp]
\begin{center}
\includegraphics[height=6cm]{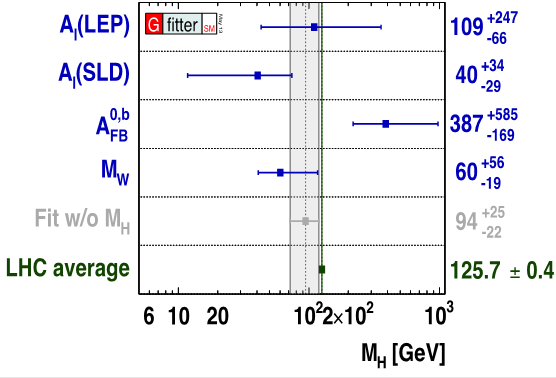}
\end{center}
\caption{Predictions from the EW Fit compared to experimental results for Higgs mass~\protect\cite{EWFit_GFitter}.}
\label{fig:EWFit2}
\end{figure}

The Higgs boson occupies a peculiar place in the Standard Model and its discovery (or proof of its nonexistence) was, after the discovery of the intermediate vector bosons 
of the weak interaction in 1983, a major milestone of experimental particle physics. In 1974, $m_H$ could span from the electron mass to $M_{Pl}$. A way to 
narrow down the mass region, and therefore help the experiments, is to use the presence of the Higgs particle in quantum corrections computed by perturbation theory,
also called 'radiative corrections'. These corrections come from virtual particles constantly produced out of 
nothing that violate the energy-conservation law by borrowing an amount of energy $E$ from the vacuum for a very short time of at most $\hbar/E$. In the electroweak sector 
these corrections are generally small compared to tree-level but still sizable. Therefore very precisely measured observables at the $Z$-pole and from $W$ mass 
measurements allow indirect constraints on undiscovered particle masses as demonstrated for the top mass in Fig.~\ref{fig:EWFit1}. Before the LHC start, 
the EW fit favored a Higgs mass around 100 GeV as shown in Fig.~\ref{fig:EWFit2}. 

%&&&&&&&&&&&&&&&&&&&&&&&&&&&&&&&&&
\subsection{LHC: collider, experiments and physics goals}
\label{sec:LHC}
%&&&&&&&&&&&&&&&&&&&&&&&&&&&&&&&&&

Following the Heisenberg uncertainty principle, $E x >\hbar c$, probing lower distance $x$ requires higher available energy $E$. For example to probe 
$x=\mathrm{O}(10^{-15})$~m, i.e. 
the proton radius, requires an energy of 0.1~GeV and therefore a moderate size apparatus. A nice illustration is provided by the charged pion of mass 0.1~GeV, 
responsible for the strong nuclear force and close to the QCD scale $\Lambda_{\mathrm{QCD}} \sim 0.2$ GeV: it was discovered in photographic plates by looking at 
atmospheric cosmic rays~\cite{Pion_DISC} and then produced by the 5~m diameter UC-Berkeley's cyclotron~\cite{Pion_Cyclo}. 

\begin{figure}[htbp]
\begin{center}    
\includegraphics[height=9cm]{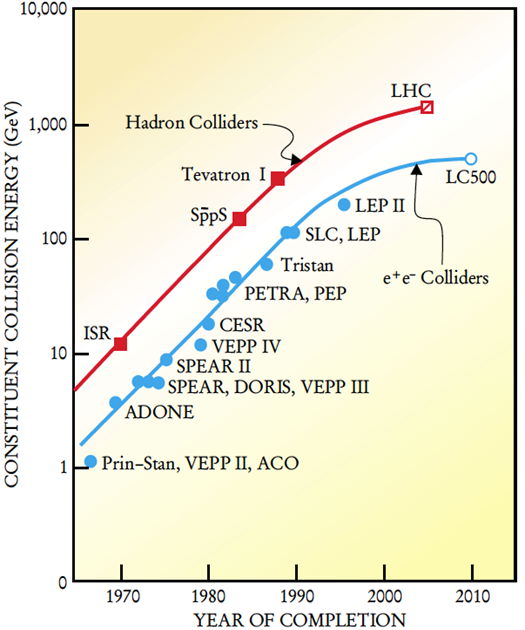}
\end{center}
\caption{Variation of the constituent energy as a function of completion time for different colliders. Constituent energy is $\sqrt{s}/6$ for 
hadronic colliders, to account for proton compositeness, and LHC refers to LHC runII~\protect\cite{LIV_ACC}.}
\label{fig:LHC1}
\end{figure}

The next interesting scale to probe is the Fermi or Weak scale ($\Lambda_{\mathrm{EW}}$) defined as:
\begin{equation}
\Lambda_{\mathrm{EW}} = {G_F}^{-1/2} \sim \mathrm{O}(10^{2})~\mathrm{GeV} 
\end{equation}
which corresponds to typical size of O(10$^{-18}$) m. Applying a linear scaling to guess the accelerator size gives a ring of O(10)~km. Giant size 
accelerator were therefore constructed to gradually reach this scale (see Fig.~\ref{fig:LHC1}). The first one to succeed was the 
proton-antiproton Sp$\bar{p}$S collider at CERN in the 1980's enabling the discovery of the $W$ and $Z$ bosons. Later e$^+$-e$^-$ LEP and SLC colliders and the proton-antiproton 
Tevatron collider at FermiLab studied these new particles in great details as discussed in Section~\ref{sec:EWFit}. However none were able 
to discover the Higgs boson because of limited statistics (Tevatron) or too low center-of-mass energy (SLC, LEP). 

\begin{figure}[htbp]
\begin{center}    
\includegraphics[height=9cm]{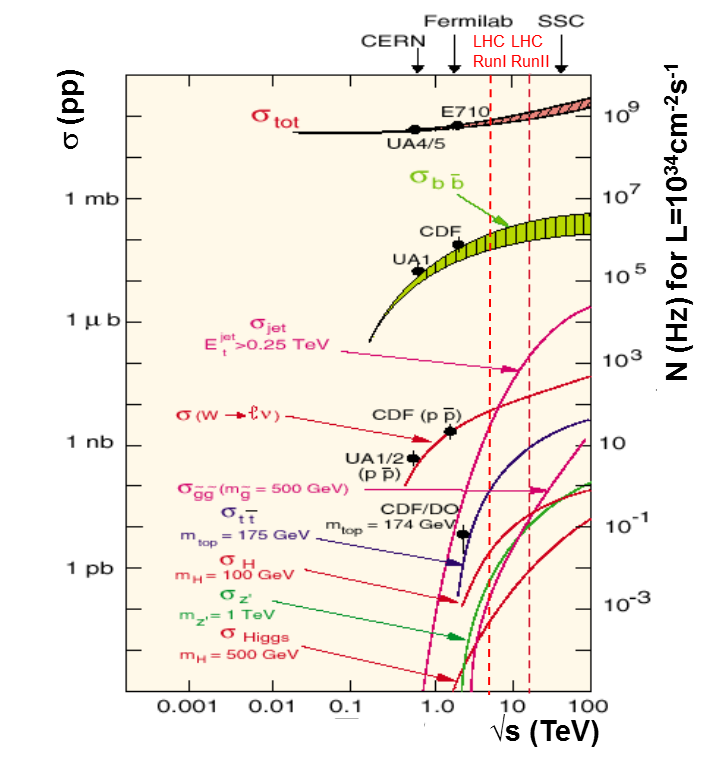}
\end{center}
\caption{Cross-section for various physics processes at different hadron colliders.}
\label{fig:LHC2}
\end{figure}

\begin{figure*}[htbp]
\begin{center}
\includegraphics[height=4cm]{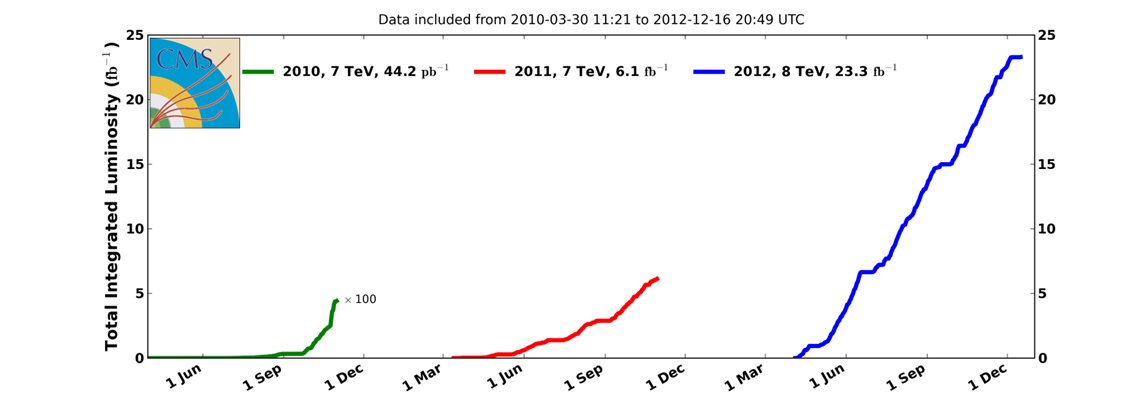}
\end{center}
\caption{LHC luminosity recorded by the CMS experiment.}
\label{fig:LHCPerf1}
\end{figure*}

\begin{figure*}[htbp]
\begin{center}
\includegraphics[height=4cm]{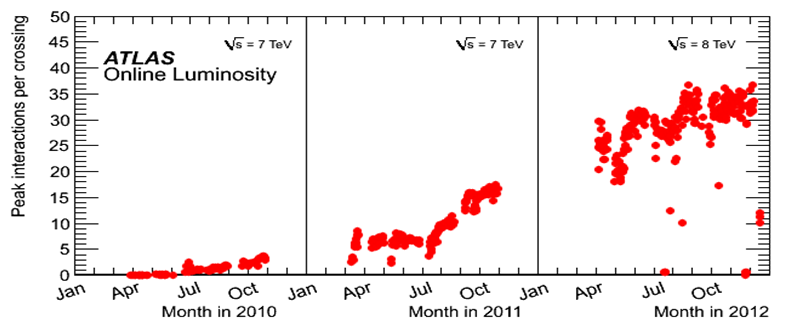}
\end{center}
\caption{Corresponding number of pile-up events as a function of time.}
\label{fig:LHCPerf2}
\end{figure*}

To cure both of these problems, it was decided in 1984 to construct a hadron collider, able to reach several times the 
center of mass energy of the Tevatron by using superconducting magnets, the only technology able to bend several TeV proton beam in the 27-km LEP tunnel.
To increase the luminosity by orders of magnitude, the collision of two proton beams (instead of proton-antiproton) is necessary and 
possible with 2-in-1 cryodipoles~\cite{LHC}. The increase of the number of protons per bunch to $10^{11}$ causes 
however an increase of the number of ``pile-up" interactions per crossing, see Fig.~\ref{fig:LHCPerf2}, that reached up to 35. This challenge,  
anticipated by the LHC experiments, leads to the construction of highly granular and radiation-hard detectors. The net result is that the LHC 
accumulated $L$ $\sim$10~fb$^{-1}$ by mid-2012, two years after starting collisions at $\sqrt{s}$=7-8 TeV, see Fig.~\ref{fig:LHCPerf1}. This 
corresponds to what the Tevatron recorded during 24 years. Thanks to these improvements in $\sqrt{s}$ and $L$, the cross-sections of interesting processes 
(bottom right part of Fig.~\ref{fig:LHC2}) are greatly enhanced. Therefore on top of being a Higgs factory, LHC offers exciting possibilities to cover 
the region 0.1-O(1) TeV where physics beyond Standard model could be hidden, see Section~\ref{sec:chap2}.

Four detectors have been built at the collision points located on the 100 m underground LEP Tunnel. Among the four,
two general-purpose experiments~\cite{ATLASDetector} and ~\cite{CMSDetector} were designed to understand the origin of the EW symmetry 
breaking mechanism, Higgs or something else, and be sensitive to any sign of new physics around the EW scale. Because of the huge complexity of 
a detector ready to cope with proton-proton collision every 25 ns and high pile-up conditions, world wide collaborations of few thousands of physicists 
and engineers were set-up, giving to these projects a 
flavor of modern cathedral dedicated to science. Note that even if this is one of the most complex and ambitious project ever built, it is comparable 
in cost to other large-size projects~\cite{BigScience}. The two detectors were based on two different technologies for the central magnets used to bend 
the charged particle trajectories: CMS uses a 4-Tesla superconducting solenoid magnet of 3m radius while ATLAS choose a smaller central solenoid 
(2 T and 1.2m radius) complemented by outer toroids~\cite{CMSvsATLAS}. These choices influence all the other detector technologies and especially 
the electromagnetic calorimeters, key to measure the kinematics of electrons and photons. CMS chose 75000 scintillating PbWO$_4$ crystals with an 
excellent energy resolution but extremely low light yield while ATLAS built a granular (200k channels) lead/liquid argon sampling calorimeter, a 
robust and well known technology, with poorer energy resolution at low energy but comparable to CMS in the ~0.1-1 TeV energy range. This complementarity 
was a key element of the Higgs discovery (Section~\ref{sec:Higgs_Disco}). Another important part of the success was the ability of ATLAS and CMS to use 
more than 90\% of the high quality data produced by LHC for physics analysis, demonstrating the excellent functioning of both experiments.

Before entering the discovery phase, the rapidly growing luminosity allowed to check in details all known SM processes and especially those related 
to the most massive particles ($W$, $Z$ and top), generally the major source of background for searches. Building on previous experiments and 
latest developments in theory computation beyond leading orders, an excellent agreement was found between predictions and data.

%&&&&&&&&&&&&&&&&&&&&&&&&&&&&&&&&&L
\subsection{Higgs and Cosmology}
\label{sec:Higgs_Cosmo}
%&&&&&&&&&&&&&&&&&&&&&&&&&&&&&&&&&

As explained before, the Higgs is considered as the cornerstone of the SM of Particle Physics. Its discovery and its properties extracted from 
the first LHC run will be first discussed and the implication of these new results for cosmology will be presented at the end of the section.

%&&&&&&&&&&&&&&&&&&&&&&&&&&&&&&&&&
\subsubsection{The Higgs discovery}
\label{sec:Higgs_Disco}
%&&&&&&&&&&&&&&&&&&&&&&&&&&&&&&&&&

\begin{figure}[htbp]
\begin{center}
\includegraphics[height=6cm]{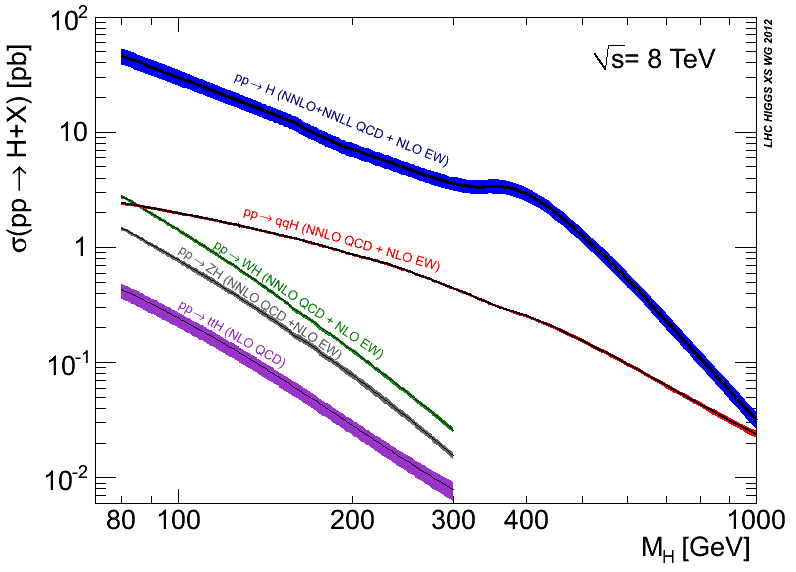}
\end{center}
\caption{SM Higgs production cross-section at LHC~\protect\cite{HXS}.}
\label{fig:HiggsLHC1}
\end{figure}

\begin{figure}[htbp]
\begin{center}
\includegraphics[height=6cm]{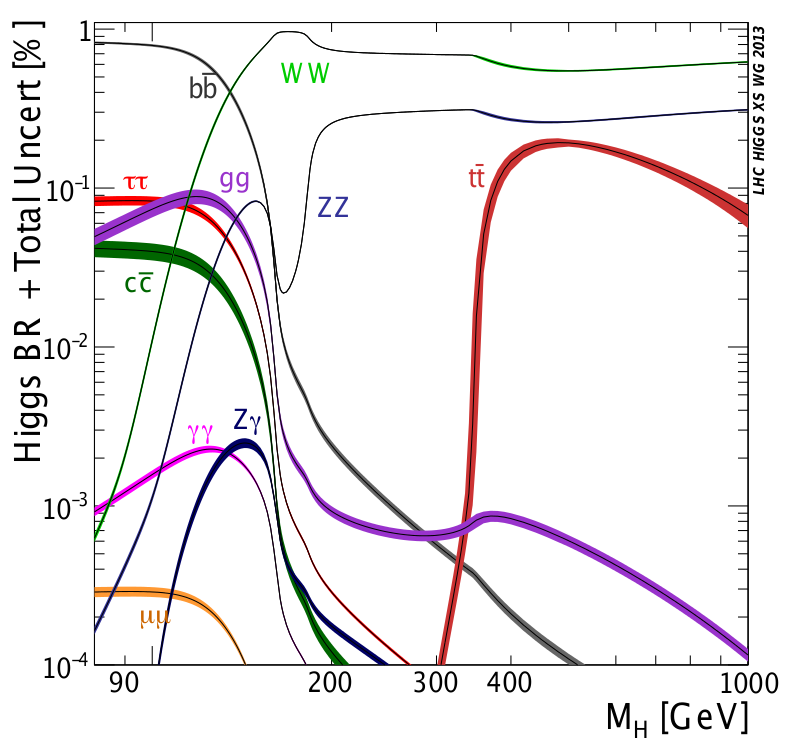}
\end{center}
\caption{SM Higgs production decay as a function of its mass~\protect\cite{HBR3}.}
\label{fig:HiggsLHC2}
\end{figure}

At LHC, the Higgs production is completely dominated by the gluon fusion process which occurs via one-loop `triangle' heavy fermion (mainly top), 
since the Higgs can not directly couples to massless gluon. Even if this process is in principle suppressed O($\alpha_S^2$), it benefits from the 
gluon domination at low momentum fraction in the proton and thus from higher $\sqrt{s}$. More precisely, the gluon fusion cross-section is multiplied 
by $\sim$25 between Tevatron and LHC run I. Sub-dominant production processes are Vector Boson Fusion (VBF) and $t\bar{t}H$ fusion, where 2 additional quarks 
(light and top quark respectively) are produced with the Higgs. Finally vector boson Higgstrahlung ($WH$, $ZH$) is also possible. Since beyond leading 
order processes over dominate the production, a huge theoretical effort was engaged to compute accurately the total cross-section for each mass 
of the Higgs boson in the range 80-1000 GeV~\cite{HXS}, see Fig.~\ref{fig:HiggsLHC1}. 
Similarly, efforts went into determining the Higgs decay branching ratios~\cite{HBR3}, as summarized in Fig.~\ref{fig:HiggsLHC2}. Before the start of the LHC, the 
theoretical uncertainties on main production and decay modes were typically 5-10\% or below. 

As already mentioned, $m_H$ was not constrained by the theory when it was introduced in the SM. 
The first effort to compute its decay branching ratios dates back from 1975~\cite{HBR1} where the range $10^{-3}<m_H<50$~GeV was considered. At that time most of 
the high mass SM particles were not discovered and therefore the results suffer from huge uncertainties. Before the LHC start, the strongest constraints were 
coming from LEP2 experiments with $m_H>114$~GeV, Tevatron experiments, $160<m_H<175$~GeV, and the unitarisation 
of the $WW$ scattering, $m_H<800$~GeV.

Therefore in 2010, the allowed mass range for the Higgs boson search was restricted to less than one order of magnitude, with a preference for the 
``low'' mass region, $\sim 100$~GeV, coming from the EW fit (Section~\ref{sec:SM}). 
The ATLAS and CMS searches were therefore divided in two categories: low mass searches where the discovery channels are 
$H\rightarrow \gamma \gamma$ (0.05 pb) and $H\rightarrow Z Z^* \rightarrow 4l$ (0.003 pb)
and high mass searches where the discovery channels are $H\rightarrow WW^{(*)} \rightarrow 2l2\nu$ (0.04 pb) and $H\rightarrow Z Z^{(*)} \rightarrow 4l, 2l2\nu$ (0.002 pb). 
In both cases $l=e,\mu$. Numbers in brackets indicate the cross-section values for $m_H=125$~GeV for low masses, and $m_H=500$ GeV for high masses.

\begin{figure}[htbp]
\begin{center}
\includegraphics[height=4cm]{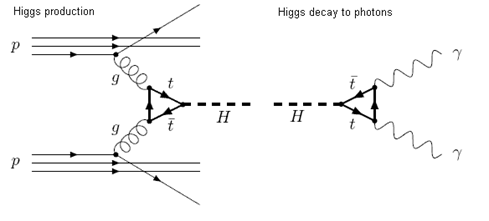}
\end{center}
\caption{Leading Feynman diagrams for the $H\rightarrow \gamma \gamma$ channel.}
\label{fig:Higgs2Photons11}
\end{figure}

\begin{figure}[htbp]
\begin{center}
\includegraphics[height=5cm]{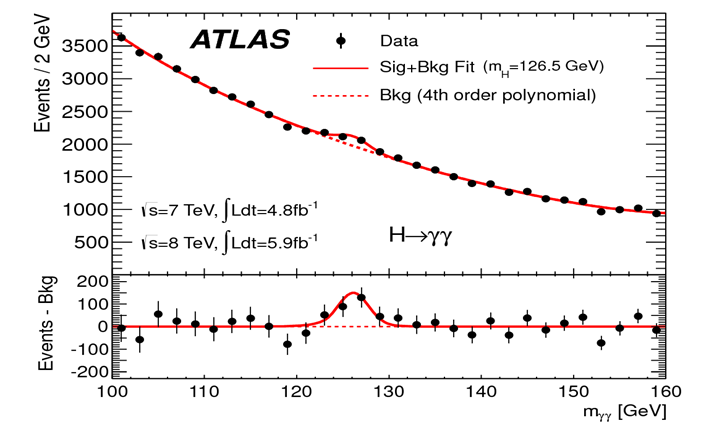}
\end{center}
\caption{Invariant mass of diphoton at the time of the discovery~\protect\cite{HIGGSDISC1}.}
\label{fig:Higgs2Photons12}
\end{figure}

As shown in Fig.~\ref{fig:Higgs2Photons11}, the $H\rightarrow \gamma \gamma$ channel can only been obtained by considering one-loop diagram in the production 
and the decay. This process, absent at tree-level, relies on the presence of virtual particles, i.e. discovering a signal in this channel will 
be a triumph for quantum field theory~\footnote{For these reasons, this channel is also sensitive to new particles in the loops.}.
In this channel, the experimental challenge consists in reducing the contribution from jet faking a photon by a factor 
$\sim$10$^4$, possible thanks to the high granular calorimeter. When achieved, the non-resonant SM $\gamma \gamma$ production becomes the dominant and irreducible 
background. A mass peak can then be searched for by fitting the side bands. With 10 fb$^{-1}$ of data, a clear peak with a significance above 
4$\sigma$ was observed at m$_{\gamma\gamma} \sim 126.5$ GeV by ATLAS, see Fig.~\ref{fig:Higgs2Photons12}~\cite{HIGGSDISC1}. 
Similar results were obtained by the CMS experiment~\cite{HIGGSDISC2}. 

\begin{figure}[htbp]
\begin{center}
\includegraphics[height=4cm]{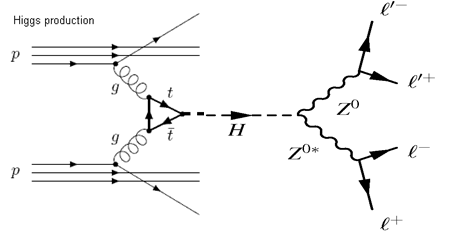}
\end{center}
\caption{Leading Feynman diagrams for the $H\rightarrow Z Z^* \rightarrow 4l$ channel.}
\label{fig:Higgs2Photons21}
\end{figure}

\begin{figure}[htbp]
\begin{center}
\includegraphics[height=7cm]{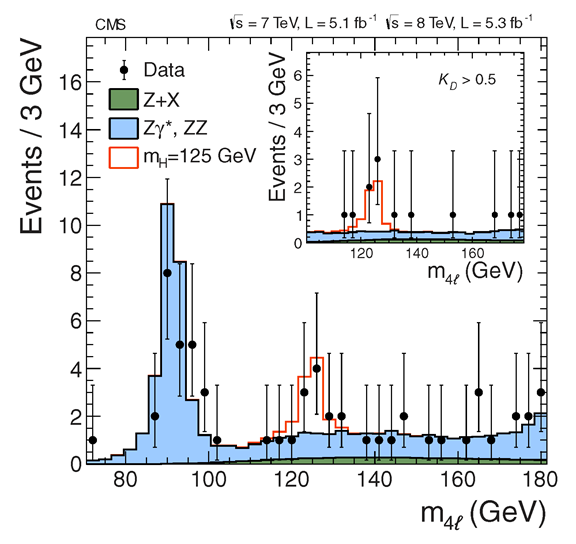}
\end{center}
\caption{Invariant mass of four-lepton at the time of the discovery~\protect\cite{HIGGSDISC2}.}
\label{fig:Higgs2Photons22}
\end{figure}

The $H\rightarrow Z Z^* \rightarrow 4l$ channel, shown in Fig.~\ref{fig:Higgs2Photons21}, is dominated by muon final states ($4\mu$ and $2e2\mu$) and therefore 
almost background free, see Fig.~\ref{fig:Higgs2Photons22}. The experimental challenge is to maximize the coverage of the four leptons and master the lepton 
energy calibration. The latter can be cross-checked with on-shell $Z\rightarrow l^+ l^-$ events, where one of the lepton radiates a photon that later converts to 
$l^+l^-$ giving a peak at 90 GeV. With 10~fb$^{-1}$ of data, a clear excess is observed over the background estimation at 125.5 GeV in CMS~\cite{HIGGSDISC2} as 
well as in ATLAS~\cite{HIGGSDISC1}. 

When combining both channels, the excess of events observed above the expected background around a mass of 125~GeV has a local significance of 5$\sigma$ for both 
ATLAS and CMS. Furthermore, the production and decay of this particle is consistent with the SM Higgs boson within uncertainties.

%&&&&&&&&&&&&&&&&&&&&&&&&&&&&&&&&&
\subsubsection{Properties of the Higgs boson}
\label{sec:Higgs_Prop}
%&&&&&&&&&&&&&&&&&&&&&&&&&&&&&&&&&

With the full run I statistics, it is already possible to study in some details the properties of this new particle. In particular, its mass can be determined at 
0.5\% precision by both experiments, i.e. more precisely than any quark mass in the SM. Current values are $125.4\pm0.4$~GeV and 
$125.0\pm0.3$~GeV for ATLAS and CMS respectively~\cite{HIGGSMASS1,HIGGSMASS2}, in agreement with the Higgs boson mass predicted by the EW fit, 
see Fig.~\ref{fig:EWFit2}. 

The spin of the new particle can be inferred from first principles and the observation of $H \rightarrow \gamma \gamma$ and $H \rightarrow ZZ^* \rightarrow 4l$ decays. 
The $\gamma \gamma$ final state forbids $J$=$n$/2 (angular momentum conservation), strongly disfavors $J$=1 since on-shell vector boson can not decay to two 
massless photons~\cite{Landau,Yang} and disfavors $J$=2 since BR($H \rightarrow \gamma \gamma$)/BR($H \rightarrow ZZ^* \rightarrow 4l$) $\sim$ 0.1 is hardly 
reproducible in graviton-inspired models. Assuming $J$=0, the parity of the new boson can be probed by looking at angular distribution of the four leptons coming 
from the Higgs decay. A negative parity is excluded at 99.6\%~\cite{HIGGSPIN} favoring $J$=0$^+$ as for the SM Higgs.

\begin{figure}[htbp]
\begin{center}
\includegraphics[height=7cm]{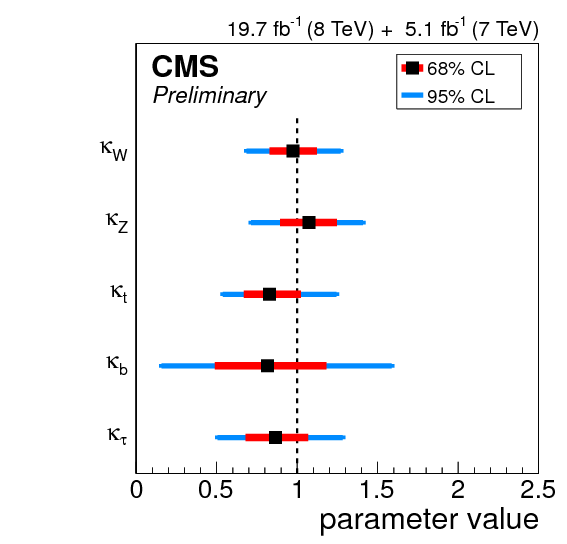}
\end{center}
\caption{Summary of the fit results for the generic five-parameter model (see text) compared to the predictions from the SM shown with dotted line~\protect\cite{HIGGSMASS2}.
Measurements are compared with the SM prediction (dotted line).}
\label{fig:HiggsProp1}
\end{figure}

\begin{figure}[htbp]
\begin{center}
\includegraphics[height=7cm]{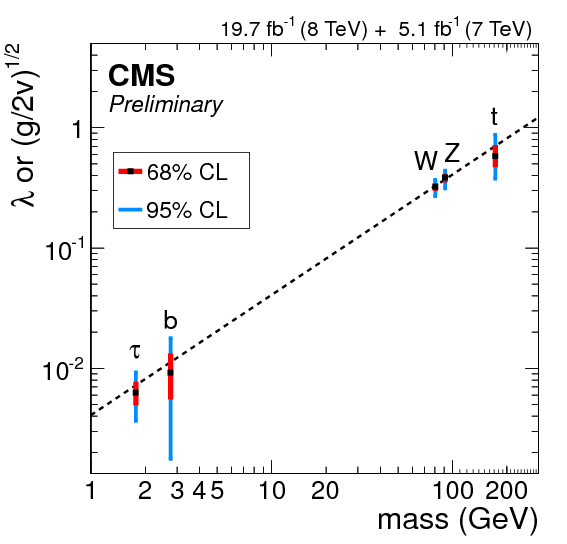}
\end{center}
\caption{Coupling strenght as a function of particle mass~\protect\cite{HIGGSMASS2}. For fermions, the coupling strenght is $\lambda=\lambda_Y$, the fitted Yukawa coupling. 
For $V$ bosons, the coupling strenght is $(g_V/(2v))^{1/2}$ where $g_V=2M_V^2/v$. Measurements are compared with the SM prediction (dotted line).}
\label{fig:HiggsProp2}
\end{figure}

Once the Higgs mass is known, all Higgs couplings can be computed within the SM. The four production modes times six decay modes from Fig.~\ref{fig:HiggsLHC1} 
and ~\ref{fig:HiggsLHC2} can be explored in principle, apart from $H\rightarrow gg, c\bar{c}$ which are not accessible because the background level is too high at LHC. 
To test the compatibility of these modes with the SM it is convenient to introduce coupling scale factors $\kappa$ defined as~\cite{HIGGSCouplings}:
\begin{multline}
\sigma \times \mathrm{BR}(jj\rightarrow H \rightarrow ii) = \\
\sigma_{\mathrm{SM}}(jj\rightarrow H) \times BR_{\mathrm{SM}}(H \rightarrow ii) \times \frac{\kappa_i^2 \kappa_j^2}{\kappa_H^2} \sim \frac{\Gamma_i \Gamma_j}{\Gamma_H}
\end{multline}
Since at LHC only the ratio of partial width can be measured (no sensitivity to $\kappa_H$ since $\Gamma_H = 4$ MeV), there are 5 relevant  
$\kappa$ parameters: $\kappa_t$, $\kappa_{\tau}$, $\kappa_{b}$ for fermions and $\kappa_W$, $\kappa_Z$ for gauge bosons. Loop induced couplings
$\kappa_{\gamma}$ and $\kappa_g$ are expressed with the previous parameters if the SM structure is assumed. Results are shown in 
Fig.~\ref{fig:HiggsProp1}~\cite{HIGGSMASS2} where a very nice agreement between the SM predictions and experimental measure can be seen, 
even if uncertainties are still large. Looking at the coupling strength as a function of the particle mass, Fig.~\ref{fig:HiggsProp2}, 
exhibits the very peculiar behavior of the SM Higgs.

All these measurements give a strong feeling that the boson discovered by ATLAS and CMS is an elementary scalar particle 
matching the Higgs boson of the Standard Model. At least this option has currently the highest probability and this will be our assumption for the remaining 
part of this section. For simplicity, we will use $m_H$=125~GeV and $\lambda=m_H^2/(2v^2)=0.13$ and also assume that there is no new physics up 
to the Planck mass. These two hypotheses (no new physics and SM Higgs boson) will be rediscussed in details in Section~\ref{sec:Higgs_NP}. To conclude 
on LHC Higgs results, it is very important to mention that using all the statistics from run I and combining $H \rightarrow ZZ \rightarrow 4l, 2l2\tau, 2l2\nu$ 
the presence of another SM Higgs-like boson is excluded below 1 TeV~\cite{HIGGSHighMass}. 

%&&&&&&&&&&&&&&&&&&&&&&&&&&&&&&&&&
\subsubsection{Higgs and the Early Universe}
\label{sec:Higgs_Universe}
%&&&&&&&&&&&&&&&&&&&&&&&&&&&&&&&&&

As already mentioned, the only way to probe experimentally the very first instant of the universe presently comes from particle physics experiments. 
However it should be said (even if it may sounds trivial) that these experiments do not reproduce all the conditions of the early universe, in particular 
there is no Hubble expansion and the matter is not in a hot plasma state with density $\rho$ and temperature $T$ connected by $\rho \propto T^4$. 
With these caveats in mind, let us illustrate the importance of the SM Higgs boson discovery on the early universe. For all these reasons, the Higgs 
mechanism occupies a central place in the early universe.

First, the Higgs mechanism provide a mass to all elementary fermions. Without this mechanism, electron will be massless and give 
macroscopic atoms, see eqn~(\ref{eq:BohrRadius}). Similarly massless quarks will prevent the atom to form since the proton can be more massive than 
the neutron (i.e. the neutron will be stable and the proton will decay to neutron, electron and neutrino). Note that $SU(2)_L \otimes U(1)_Y$ will 
still be broken at $\Lambda_{\mathrm{QCD}} \sim 0.2$ GeV and will give massive gauge bosons of 30 MeV, i.e. weak interaction will be strong~\cite{NoEWSB}.

\begin{figure}[htbp]
\begin{center}
\includegraphics[height=5cm]{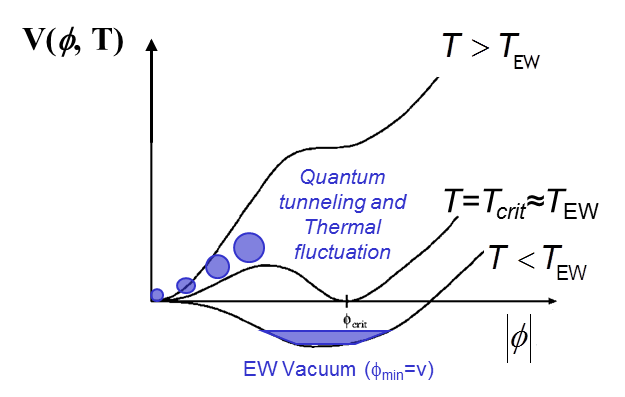}
\end{center}
\caption{Configurations of the Higgs potential with $m_H<$ 75 GeV.}
\label{fig:HiggsPotential1}
\end{figure}

\begin{figure}[htbp]
\begin{center}
\includegraphics[height=5cm]{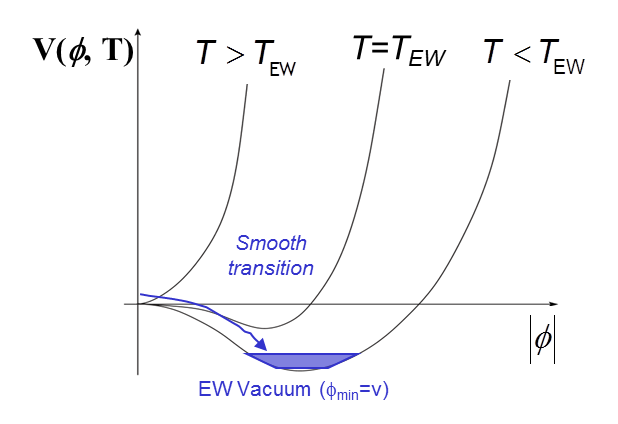}
\end{center}
\caption{Configurations of the Higgs potential with $m_H>$ 75 GeV.}
\label{fig:HiggsPotential2}
\end{figure}

Second, the form of the Higgs potential and the exact value of $m_H$ could have dramatically changed the universe 
that we know. The high temperature expansion of the Higgs potential at one-loop level can be written to a constant~\cite{CosmoRef}:
\begin{equation}
V(\phi,T) = a_T \phi^2 + b_T \phi^3 + c_T \phi^4
\end{equation}
The $a_T$, $b_T$ and $c_T$ parameters are defined as ($m_t^4 \gg m_W^4, m_Z^4$ is assumed)
\begin{equation}
\begin{array}{lrc}
a_T = a(T^2 - T_{\mathrm{EW}}^2) \simeq \frac{m_{tWZ}^2}{4v^2}(T^2 - T_{\mathrm{EW}}^2) \simeq 0.33(T^2 - T_{\mathrm{EW}}^2) \\
b_T = bT = -\frac{m_Z^3 + 2 m_W^3}{\sqrt{2} \pi v^3}T \simeq -0.027 T \\
c_T = \frac{m_H^2}{2v^2} - \frac{3m_t^4}{2\pi^2v^4}\mathrm{ln}\frac{T}{m_t} \simeq \frac{m_H^2}{2v^2} \simeq \lambda \simeq 0.13 \\
\end{array}
\label{eq:paramHiggs}
\end{equation}
where $m_{tWZ}^2=2m_t^2+2m_W^2+m_Z^2\sim (285)^2~\mathrm{GeV^2}$. It is interesting to note that, for a fixed temperature, the numerical values of 
these parameters depends only on $W$, $Z$, top and Higgs masses. The phase transition will occur at $T_{\mathrm{EW}}$:
\begin{multline}
T_{\mathrm{EW}}^2= \frac{1}{2a} \left(m_H^2 + \frac{3 m_t^4}{2 \pi^2 v^2} \right) = \frac{1}{2a} \left(m_H^2 + \Delta m_H^2 \right) \\
\simeq 1.5 \left(m_H^2 + (47 \mathrm{GeV})^2 \right)
\end{multline}
Note that neglecting the temperature and the radiative correction terms $a_T \rightarrow -aT_{\mathrm{EW}}^2=-m_H^2/2$, $b_T \rightarrow 0$ and $c_T \rightarrow \lambda$ 
giving back eqn~(\ref{eq:HiggsPotential}). For $T>T_{\mathrm{EW}}$, the universe is symmetric around $\phi_{min}=0$ since $a_T>0$, and SM particles are massless. At the phase 
transition, $T=T_{\mathrm{EW}}$, $a_T=0$ and the potential shape depends only on $b$ and $\lambda$. If $|b|\geq \lambda/\sqrt{3}$, i.e $m_H<75$ GeV, a second minimum of the 
potential appears at some critical temperature close to $T_{\mathrm{EW}}$ and its depth is equivalent to the minimum at $\phi_{min}=0$. A first order transition could then 
start by quantum tunneling, see Fig.~\ref{fig:HiggsPotential1}. For long, it was a nice explanation for baryogenesis, called EW baryogenesis, 
since CP violation on bubble surface could froze due to the phase transition~\cite{EWBaryo1,EWBaryo2,EWbaryo3} but this case is now excluded by the Higgs discovery at 
$m_H=125$~GeV. In contrast we are in the case $|b|<\lambda/\sqrt{3}$, i.e $m_H>75$ GeV, where the transition between the two minima is smooth and play no role in the 
baryogenesis. The term $\phi^3$ can be neglected and the only minimum of the potential, at tree level and low temperature, is located at 
$\phi_{min} = (aT_{\mathrm{EW}}^2/\lambda)^{1/2}=v$, as shown in Fig.~\ref{fig:HiggsPotential2}. 

Third, the discovery of an elementary scalar field is of primordial importance. The cosmological observations of recent decades revealed that the universe expanded with 
acceleration on two different stages of its evolution: in the very beginning and at present time. The former is presently
described by the theory of inflation~\cite{Inflation1,Inflation2,Inflation3,Inflation4}, the cornerstone of the Standard Model of cosmology. In this 
short period of time (t=10$^{-36}$-10$^{-33}$s), the matter
content of the universe must be dominated by a fluid with negative pressure, described by quantum field theory (given the high energy). The prototype is a
scalar field $\phi$, called inflaton, since this is the only possibility compatible with symmetries implied by the cosmological principle. It is therefore
very natural to ask whether the Higgs can play the role of the inflaton. For this to happen, it is mandatory to add a term $\xi \phi^{\dag} \phi R$ to the 
Lagrangian~\cite{HiggsGravity}, coupling non minimally the Higgs field to gravity, where $R$ is the Ricci scalar and $\xi$ is a dimensionless 
coupling constant generally large, O(10$^4$). In this context, the dimensionless quantity $\Psi=\sqrt{\xi}\phi/M_{Pl}$ can control the cosmology:
\begin{itemize}
\item $\Psi \gg 1$ implies $\phi \gg M_{Pl}/\sqrt{\xi}$, and the slow-roll inflation takes place.
\item $\Psi \sim 1$ implies $\phi = M_{Pl}/\sqrt{\xi}$, inflation ends and $T_R \geq 10^{13}$ GeV.
\item $\Psi \ll 1$ implies $\phi \ll M_{Pl}/\sqrt{\xi}$ gets a vev as discussed earlier.
\end{itemize}
This is a seductive approach able to make predictions at two-loop level~\cite{HiggsInflaton3,HiggsInflaton4,HiggsInflaton5}. Those predictions can be compared to precise 
experimental cosmological and particle physics measurements. For example, Higgs inflaton predictions match perfectly the latest Planck data~\cite{HiggsInflaton_Planck}. 
It is however fair to mention that there is still some debate in the theory community about its validity~\cite{HiggsInflatonPb}.

Let us conclude this section by briefly mentioning the link between the Higgs boson and the cosmological constant. Latest Planck results ($\Omega_{vac}\sim0.7$) 
confirmed that acceleration is taking place at present. It is therefore tempting to identify the constant energy density of the vacuum ($\rho_{vac}$) 
to the EW vacuum. Plugging the vacuum configuration $<\phi>=\frac{1}{\sqrt{2}}\binom{0}{v}$ in eqn~(\ref{eq:HiggsPotential}) gives, 
$\rho_{vac} = -m_H^2 v^2/8 \sim - 10^8~\mathrm{GeV}^4$, highly incompatible with 
the cosmology measurements $\rho_{vac} = \Omega_{vac} \rho_{cri} \sim 10^{-48}~\mathrm{GeV}^4$. This is the cosmological constant problem~\cite{CCPb1}. For recent 
updates, see ~\cite{CCPb2,CCPb3}.

%&&&&&&&&&&&&&&&&&&&&&&&&&&&&&&&&&
%&&&&&&&&&&&&&&&&&&&&&&&&&&&&&&&&&
\subsection{The Higgs discovery calls for New Physics?}
\label{sec:Higgs_NP}
%&&&&&&&&&&&&&&&&&&&&&&&&&&&&&&&&&

The present situation in Particle Physics is paradoxical. On one hand, the Standard Model is amazing successful 
to describe all experimental data and since 40 years no significant deviation has been detected. 
On the other hand, it is plagued by many theoretical problems: very different mass and mixing for the 12 fermions (flavor problem), fine-tuned parameters 
($m_H$, see below, and $\Theta_{QCD}$) and no GUT-scale unification of forces. Moreover the SM does not include gravity 
and consequently can not describe any large scale or very high energy phenomena. Finally, it does not have any good dark matter candidates (Section~\ref{sec:DM}). 
Therefore the Standard Model can not be the ultimate theory and should be valid only up to a certain scale ($\Lambda_{NP}<M_{Pl}$).

\begin{figure}[htbp]
\begin{center}
\includegraphics[height=5cm]{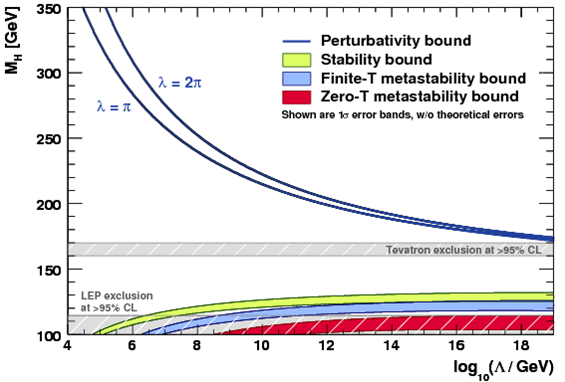}
\end{center}
\caption{Stability of the EW vacuum as a function of $\Lambda_{NP}$ before the Higgs discovery~\protect\cite{HiggsNewPhysics1}.}
\label{fig:HiggsNewPhysics1}
\end{figure}

\begin{figure}[htbp]
\begin{center}
\includegraphics[height=5cm]{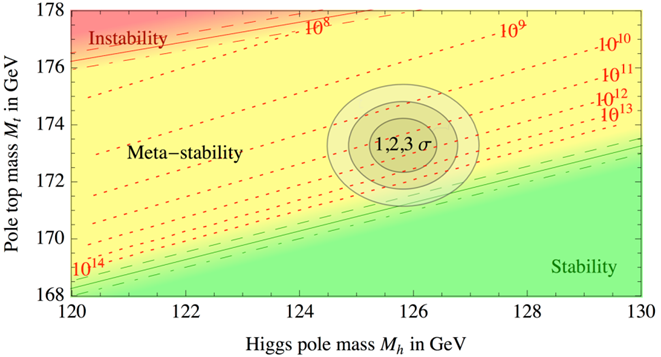}
\end{center}
\caption{Stability of the EW vacuum as a function of $\Lambda_{NP}$ after the Higgs discovery~\protect\cite{HiggsNewPhysics2}.}
\label{fig:HiggsNewPhysics2}
\end{figure}

Before going further, it is interesting to check whether the Higgs potential could survive up to $M_{Pl}$ in the absence of any new physics. 
For very high energies, the Higgs potential can be approximate to $V = \lambda \phi^4$, i.e. the potential will only depend on the Higgs self coupling $\lambda$.
If $\lambda\sim0$, the EW vacuum is metastable and could become unstable if $\lambda<0$ which gives the lower bounds of 
Fig.~\ref{fig:HiggsNewPhysics1},  as of 2009~\cite{HiggsNewPhysics1}. Today we know that $m_H\simeq 125$~GeV, which means that the instability bound
is crossed at $\Lambda \sim 10^{10}$ GeV. This is may be a hint on the presence of new physics at or before this scale. Pushing further to the Planck 
scale gives an intriguing result: the vacuum is metastable at 2$\sigma$ level, i.e. its lifetime is greater than the age of the universe, as 
illustrated in Fig.~\ref{fig:HiggsNewPhysics2}~\cite{HiggsNewPhysics2}. It is noteworthy that the result 
is now driven by the top mass precision.

Despite this intriguing result, solving the long list of SM problems discussed above requires the introduction of new particles, generally predicted by theories 
beyond the Standard Model (BSM). The most severe problem is coming from the fine-tuning of the Higgs mass and is called the hierarchy or 
naturalness problem~\cite{Hierarchy_pb1,Hierarchy_pb2}. 
Hierarchy problem because one has to explain the extreme weakness of the gravity at short distance, reflected in the ratio between the Fermi and the Newton constants
$(G_F \hbar^2)/(G_N c^2) \sim 1.7 \times 10^{33}$. It can be reformulated in terms of the Higgs mass divergence at high energy since $G_F \propto m_H^{-2}$, see 
eqn~(\ref{eq:HiggsVev}). Here, the problem is that an elementary spin-0 particle mass is not protected by any symmetry unlike massive spin-1 and spin-1/2 particles 
which are protected by broken gauge and chiral symmetries, respectively. As a massive scalar couples to all 
virtual particles present in the vacuum with an energy $E$ ($\Lambda_{NP} = \mathrm{max} (E)$), its radiative corrections can be expressed as~\cite{Giudice_Natural}:
\begin{multline}
\Delta m_H^2 = \frac{3G_F}{4\sqrt{2} \pi^2} (4 m_t^2 -2 m_W^2 -m_Z^2 - m_H^2) \Lambda_{NP}^2 = \\
\kappa \Lambda_{NP}^2 \simeq 0.05 \Lambda_{NP}^2
\label{eq:kappa}
\end{multline}
This is the naturalness problem: $m_H^2$ requires a very high adjustment between the bare mass of the 
Higgs, $(m_H)_0^2$ and $\Delta m_H^2$. For $\Lambda_{NP}=M_{Pl}$ this will look like: $m_H^2 = $ 
36127890984789307394520932878928933023 - \\
36127890984789307394520932878928917398.
 
To restore naturalness, three main possibilities have emerged:
\begin{itemize}
\item Add a new broken symmetry between fermion and boson, called supersymmetry (SUSY), which protect the Higgs mass with new weakly coupled particles;
\item Assume that Higgs is not elementary but a strongly coupled composite particle, requiring a reformulation of the problem;
\item Assume extra spatial dimensions where gravity propagates in, which explains the weakness of gravity in our 4D world. 
\end{itemize}
All these BSM theories predict new particles, partners of $t$, $W$, $Z$ and sometimes Higgs, below few TeV to dump the Higgs mass 
quadratic divergence. Generally a new conserved quantum number (parity) is attached to these new particles offering good candidate for dark 
matter~\cite{DMCandidates2}. Moreover, assuming that naturalness is a guiding principle of Nature, which worked amazingly well in particle physics 
during the XX$^{\mathrm{th}}$century~\cite{Giudice_Natural,Murayama_SUSY}, it is possible to guess $\Lambda_{NP}$. For example requiring
$\Delta m_H^2 < m_H^2$ gives $\Lambda_{NP} < 550$ GeV, i.e. new physics directly accessible at LHC. Aside from a direct detection, 
the presence of new physics close to the EW scale could also cause observable deviations 
in all EW precision measurements (including now Higgs couplings) with respect to the SM predictions. 

For all these reasons, signatures predicted by ``natural" theories are guiding the new physics searches at LHC and 
are discussed in details in Section~\ref{sec:chap2}. Of course other attempts have been made to build BSM theories regardless of the
naturalness argument. For example starting from the SM particle structure, it is reasonable to expect other quark and lepton generations and/or 
extra bosons ($Z'$, $W'$, $g'$) as well as other Higgs bosons ($n>1$ doublets instead of the minimal $n=1$ SM solution, or even 
triplets). Right-handed neutrinos, new symmetry between bosons and fermions or leptons and quarks could also be imagined. 
Pushing even further, asymmetric left and right-handed chiral multiplets could be part of a more general multiplet at GUT scale. Finally in order to 
reduce the number of SM parameters, a high energy unification of forces could be envisaged and/or mechanisms could be developed to explain 
SM parameter values and in particular the quantized value of electric charge giving $Q(e)+2Q(u)+Q(d)=Q(u)+2Q(d)=\mathrm{0}(10^{-21})$. Natural theories generally 
integrate or were build upon these ideas. Most of these SM extensions can be seriously challenged by LHC results as foreseen before LHC start~\cite{Plehn_BSM}. 
Section~\ref{sec:chap2} presents now in detail the status of these theories in light of LHC run I results.

%&&&&&&&&&&&&&&&&&&&&&&&&&&&&&&&&&&&&&&&&&&&&&&&&&&&&&&&&&&&&&&&&&&&&&&&&&&&&&&&&&&&&&&&&&&&&&&&&&&&&&&&&&&&&&&&&&&&&&&&&&&&&&&&&&&&&--------------------------------------------------
%&&&&&&&&&&&&&&&&&&&&&&&&&&&&&&&&&
%&&&&&&&&&&&&&&&&&&&&&&&&&&&&&&&&&
%&&&&&&&&&&&&&&&&&&&&&&&&&&&&&&&&&
\newpage
\section{Beyond Standard Model and Cosmology}
\label{sec:chap2}
%&&&&&&&&&&&&&&&&&&&&&&&&&&&&&&&&&

LHC gives an unique opportunity to explore the uncharted 0.1-O(1) TeV territory with 20 fb$^{-1}$ of data at $\sqrt{s}$=8 TeV, where 
first hints from BSM physics are expected. 
This section presents in details the ``eagerly awaited" results from the direct searches of BSM particles and discusses the impact on natural 
theories: SUSY in Section~\ref{sec:BSM_SUSY}, extra dimensions and composite Higgs in Section~\ref{sec:BSM_NonSUSY}. Other BSM
models are briefly mentioned in Section~\ref{sec:OthersBSM_noHierarchyPb}. The consequences of these searches on dark matter are presented in  
Section~\ref{sec:DM}. Apart from LHC results probing the energy frontier, other areas of particle physics could provide evidence 
for BSM physics. This is the case of the neutrino sector, briefly reviewed in Section~\ref{sec:neutr}, which could give an elegant solution to 
matter-antimatter asymmetry in the early universe. This section ends by presenting future prospects for particle physics in Section~\ref{sec:prosp}.

%&&&&&&&&&&&&&&&&&&&&&&&&&&&&&&&&&
\subsection{Supersymmetry searches at LHC}
\label{sec:BSM_SUSY}
%&&&&&&&&&&&&&&&&&&&&&&&&&&&&&&&&&

Supersymmetry is the leading theory for physics beyond the Standard Model since it provides a solution
to most of its shortcomings, including the hierarchy problem, and is based on very solid foundations~\cite{SUSY}. It is the accomplishment of theoretical efforts 
started in the 70's and aiming at symmetrizing boson and fermion fields. This is done by creating chiral superfields, composed of a complex scalar and a spinor fermion, 
which transform in superspace via a new symmetry of space-time called supersymmetry. 
It was demonstrated that SUSY is the only non trivial extension of the Poincar\'e group, the space-time symmetry of the Standard 
Model. It was also realized that the commutators of two local SUSY transformations give a local translation: therefore 
local SUSY naturally implies gravity with two gauge fields, the graviton $G$ and the gravitino $\tilde{G}$, a step forward compare to the SM, which 
has no description of gravity. The SUSY phase space is huge and only the part can be probed at LHC is mentionned hereafter. 

%-----------------------------------------------------------------------
\subsubsection{Minimal SuperSymmetric Model (MSSM) and natural spectrum}
\label{sec:SUSY_MSSM}

\begin{figure}[htbp]
\begin{center}
\includegraphics[height=6cm]{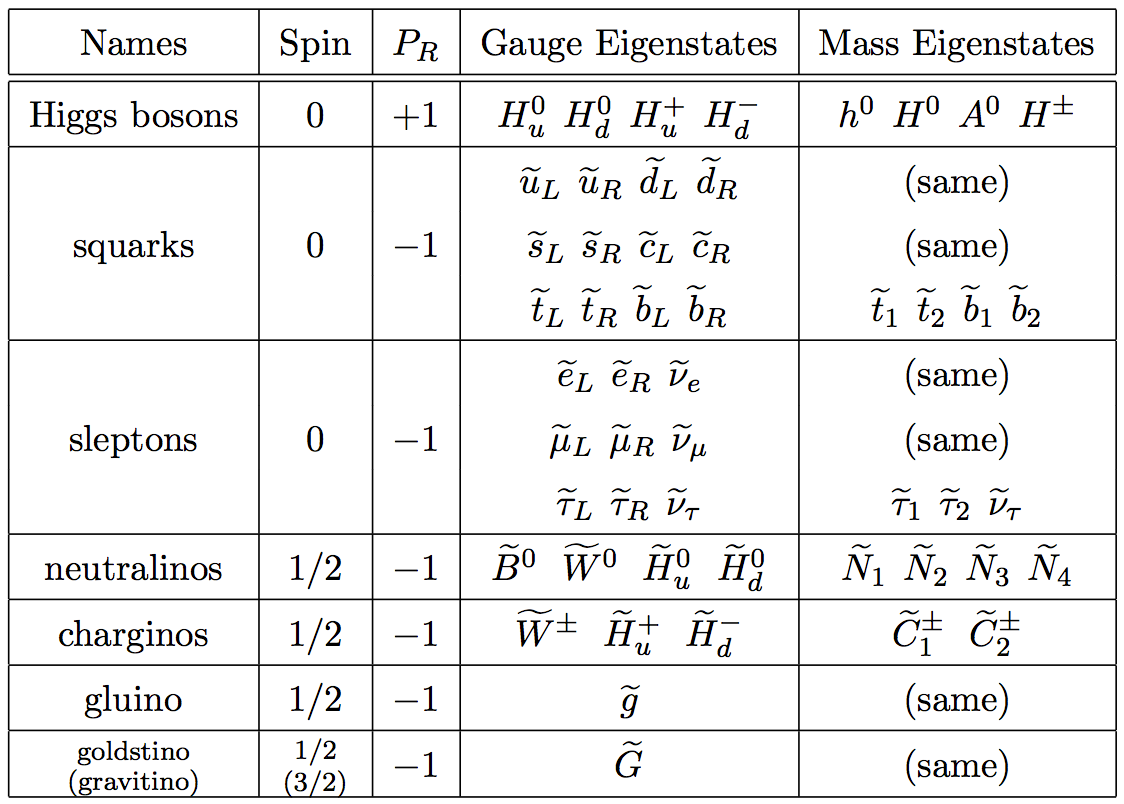}
\end{center}
\caption{SUSY particles predicted by MSSM~\protect\cite{MSSM}.}
\label{fig:SUSYGene1}
\end{figure}

After a decade of theoretical work a realistic SUSY model, in form of the minimal SM extension that realizes N=1 supersymmetry, 
was proposed at the beginning of the 80's~\cite{MSSM}. This model predicts new particles, called sparticles, that are the superpartner  
of each SM particle in the chiral multiplets. A new quantum number, R-Parity, negative/positive for SUSY/SM particles, is created. 
The sparticles have therefore the same quantum numbers as their SM partners, except for the spin, half a unit of spin different, and R-Parity.
A remarkable by-product is that these new particles allow for the unification of forces  
at the GUT scale, solving one the SM problem. The complete list of sparticles is given in Fig.~\ref{fig:SUSYGene1} and detailed briefly below. 
 
To give masses to up and down-type fermions, the SM Higgs sector needs to be extended by adding another $SU(2)_L$ complex doublet. As a result, 8 mass eigenstates 
exists: three neutral Higgses ($h^0$, the lighter, $H^0$ and $A^0$), two charged  Higgses ($H^{\pm}$) and three Goldstone bosons 
($G^0$, $G^{\pm}$) 'eaten' to give $W$ and $Z$ masses. Each of the neutral component of the doublet acquires a vev called $v_u$ and $v_d$ related by 
$v_u^2+v_d^2= v^2$ and $\mathrm{tan}\beta=v_u/v_d$, to get the correct $W$ and $Z$ masses.

The other new particles are the squarks $\tilde{q}$ and the 
sleptons $\tilde{l}$, spin-0 partners of the SM fermions. Similarly, Wino $\tilde{W}$, Bino $\tilde{B}$ and Higgsinos 
$\tilde{H}_{u,d}^{0,\pm}$ are the spin-1/2 superpartners of the EW bosons and mix to give EWKinos decomposed in 4 neutralinos $\tilde{\chi}_{1,2,3,4}^0$ (noted $\tilde{N}$ in Fig.~\ref{fig:SUSYGene1})
and 4 charginos $\tilde{\chi}_{1,2}^{\pm}$ (noted $\tilde{C}$ in Fig.~\ref{fig:SUSYGene1}). To complete the list, colored gluinos $\tilde{g}$ and the gravitino $\tilde{G}$ are the partners 
of the gluon and graviton. Note also that left and right-handed fermions have two different SUSY partners $\tilde{f}_{L,R}$ that could mix to give mass eigenstates 
$\tilde{f}_{1,2}$ provided the SM partner is heavy enough, like in the third generation. With this set-up the number of fermions and bosons is equalized and among others 21 new 
elementary scalar particles are predicted. This generally explains the lower SUSY production cross-section compare to other BSM theories and makes non-colored sparticle 
discovery particularly challenging. Since each SM particle and its superpartner belong to the same multiplet, the sparticle decay generally 
involves the SM partner. However due to the high number of new particles many different decays are possible depending on the sparticle 
mass spectrum. This other reason explains also why the sparticle discovery is an extremely challenging task.
 
\begin{figure}[htbp]
\begin{center}
\includegraphics[height=7cm]{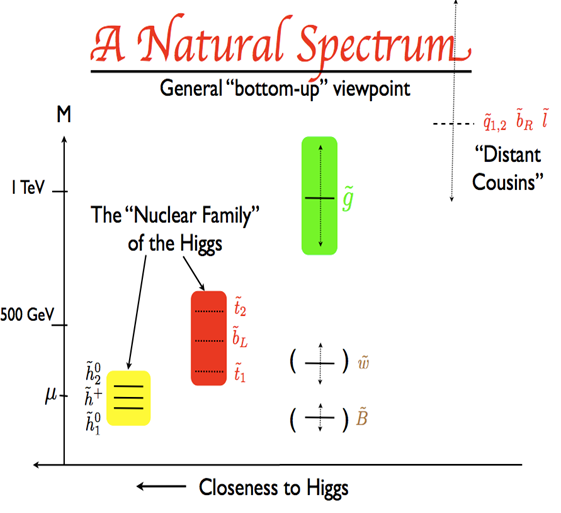}
\end{center}
\caption{Natural SUSY particle mass spectra giving less than 10\% tuning~\protect\cite{SUSY_NatSpectrum}.}
\label{fig:SUSYGene2}
\end{figure}

Even if it is not the only possible realization of SUSY, MSSM still serves as a reference for today's searches since it provides a very elegant 
solution to the hierarchy problem: if in each SUSY multiplet sparticle and particle have the same mass, the coupling of sparticles with the Higgs 
removes exactly the quadratic mass divergence, 
i.e $\kappa=0$ in eqn~(\ref{eq:kappa}). Even if the mass degeneracy is disproved experimentally, the introduction of sparticles replaces the quadratic 
divergence by a logarithmic divergence of the form $(m_{\tilde{f}}^2-m_{f}^2) \mathrm{ln}(\Lambda_{NP}/m_H)$, a substantial gain for naturalness. 
Imposing a natural theory will therefore imply that all heavy particles entering eqn~(\ref{eq:kappa}) have masses close to their SM partners: 
stop should be close to the top quark mass, Wino and Bino close to $W$ and $Z$ masses and Higgsinos, governed by the $\mu$ parameter 
appearing at tree level, very close to the Higgs mass.
Since stop is a scalar, its mass will quadratically diverge unless it is protected by a O(TeV) gluino appearing in the loop 
$\tilde{t}_1 \rightarrow \tilde{g} t \rightarrow \tilde{t}_2$. Since the left-handed bottom is part of the SM doublet including left-handed top, 
the left-handed sbottom should also be light. Finally because of the lower Yukawa couplings of leptons and other quarks, sleptons and other squarks are less constrained. 
They are even required to be heavy and degenerate to avoid too high CP violation and/or Flavor Changing Neutral Current (FCNC) already excluded experimentally. 
Figure~\ref{fig:SUSYGene2} shows natural SUSY particle mass spectra integrating these constraints and giving less than 10\% tuning. As before 2010 no exploration 
of this natural spectrum was possible, the LHC experiments were ideally placed to discover or disprove the presence of these new particles.

The other consequence of the non-mass degeneracy of particle and sparticles is that SUSY should be broken. However, unlike for the EW symmetry, it was realized at 
the end of 70's that SUSY can not be spontaneously broken. Instead it is softly broken in a hidden sector that communicates to the 
visible sector via a messenger that could be gravity (supergravity (SUGRA)-like models) or gauge bosons (gauge-mediated SUSY-breaking (GMSB)-like models). 
For gravity mediation an alternative is that no tree-level coupling transmits the SUSY breaking and sparticles masses are generated by one or two loop diagrams 
(anomaly-mediated SUSY breaking (AMSB)-like models). In any case, the price to 
pay for the soft SUSY breaking is the addition of 105 new parameters compared to the SM. 

%-----------------------------------------------------------------------
\subsubsection{Search strategy at LHC}
\label{sec:SUSY_Search}

ATLAS and CMS have developed a rich and coherent program to discover SUSY particles which resulted in about 200 public analyses~\cite{ATLAS_SUSY_Res,CMS_SUSY_Res}. 
The program mainly focuses on models where R-parity is conserved (RPC) since this is a simple way to prevent a too fast proton decay, and will provide 
a very good candidate for dark matter. This assumption has two important phenomenological consequences: first the lightest SUSY particle (LSP) is 
stable and corresponds to the massive lightest neutralino ($\tilde{\chi}_1^0$) or the approximately massless gravitino ($\tilde{G}$) in SUGRA-like or GMSB-like
models, respectively. Second, sparticles will be pair-produced at LHC. Highest cross-sections are expected from gluino-gluino, squark-antisquark and gluino-squark production 
(strong SUSY): for a 1 TeV gluino, $\sigma(\tilde{g}\tilde{g})=0.05$ pb giving 1000 events at LHC run I. This is typically 10 times the first and second generation 
squark-antisquark cross-section and 100 times the stop-antistop cross-section. EWKinos production cross-sections are much lower than strong SUSY. 
A 400 GeV pair-produced chargino has also $\sigma=0.05$ pb, typically 100 times more than pair-produced leptons of the same mass.

To clarify the presentation of the results, RPC searches are shown in sequence: $i)$ gluinos, first and second generation squarks, $ii)$ third generation squarks, $iii)$ 
EWKinos and sleptons $iv)$ a summary of RPC searches after run I. Searches for other signatures including R-Parity Violated (RPV), Long-lived particles or 
beyond MSSM solutions are then discussed. For conciseness and pedagogic reasons, only ATLAS results 
are reported since CMS obtained very similar results. A full ATLAS and CMS review can be found in~\cite{SUSY_Ref3}. To be complete SUSY Higgs searches 
are first recalled. All limits below are reported at 95\% Confidence Level (CL). 

%-----------------------------------------------------------------------
\subsubsection{SUSY Higgs searches}
\label{sec:SUSY_H}

The newly discovered Higgs boson may well be the lightest neutral Higgs of the MSSM ($h^0$) since 
it possesses very similar properties as the SM one when $m_A^2 \gg m_Z^2$ and $\tan \beta>1$ (decoupling limit). Note, however, that 125 GeV is close to the upper mass bound 
of possible lightest Higgs masses in MSSM and requires high stop masses, in tension with the natural SUSY spectrum. Extra neutral and charged Higgses, which 
preferentially couple to the most massive down-type 
fermions, are also actively searched. At the LHC, neutral Higgses are produced singly or accompanied by $b$-jet(s) and decay 
via $\tau^+ \tau^-$, $b\bar{b}$ and more marginally $\mu^+ \mu^-$. Charged Higgses with lower masses than the top quark 
predominantly appear in top decays $t \rightarrow b H^{\pm} \rightarrow b\tau \nu$. On the other hand, charged Higgses with higher masses than the top quark are  
produced in association with top and bottom quarks and decay via $H^{\pm} \rightarrow t b$. 

Searches therefore focus on final states with $t$, $b$ and/or $\tau$. Most up-to-date searches for neutral Higgses~\cite{MSSM_CMS_phi1,MSSM_CMS_phi2}, 
as well as for charged Higgses~\cite{MSSM_ATLAS_H+} are interpreted in the ($\mathrm{tan}\beta$, $m_A$) plane for neutral Higgses and ($\mathrm{tan}\beta$, $m_{H^{\pm}}$) plane for 
charged Higgses, 
there being the relevant tree-level parameters since $m_{H^{\pm}}^2=m_A^2+m_W^2$. Other SUSY parameters, entering via radiative corrections, are 
fixed to particular benchmark values, chosen to exhibit certain MSSM features. The most commonly used scenario, called $m_h^{max}$~\cite{MSSM_Higgsmodel}, 
maximizes the lightest neutral Higgs mass for a fixed $\mathrm{tan}\beta$ and large $m_A$, while the stop and sbottom masses are around 1 TeV.
In this scenario, the null result on neutral Higgs searches can rule out models with $m_A < 125$ GeV as well as large values of tan$\beta$ ($> 5$). For $m_{H^{\pm}} < m_t$,
$H^{\pm}$ masses are practically excluded below 160 GeV for all $\mathrm{tan}\beta$ values while if $m_{H^{\pm}} > m_t$, most of the ($\mathrm{tan}\beta$, $m_{H^{\pm}}$) plane
is still not excluded. Those results favor neutral and charged SUSY Higgses with masses higher than $h^0$ mass, even if no model-independent limits yet exist.

%-----------------------------------------------------------------------
\subsubsection{Direct searches of gluinos and first- and second-generation squarks}
\label{sec:SUSY_gluino}

\begin{figure}[htbp]
\begin{center}
\includegraphics[height=6cm]{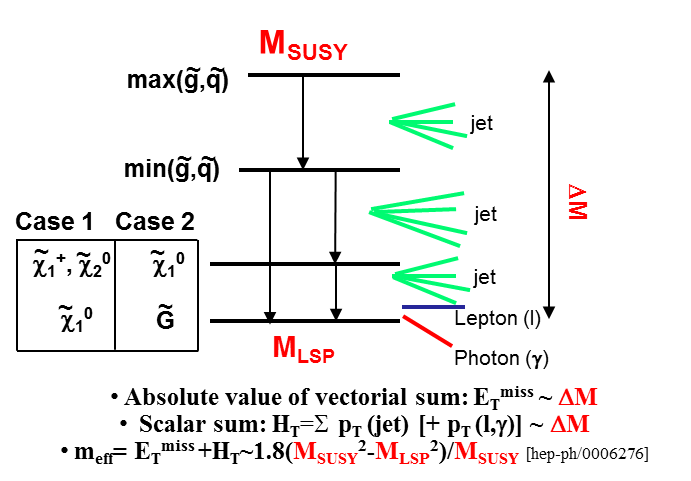}
\end{center}
\caption{Strategy of the ATLAS search for colored SUSY particles. The two cases shown in at the bottom of the SUSY spectrum correspond to the two considered LSP types~\protect\cite{SUSY_Ref3}.}
\label{fig:SUSYStrategy1}
\end{figure}

\begin{figure}[htbp]
\begin{center}
\includegraphics[height=7cm]{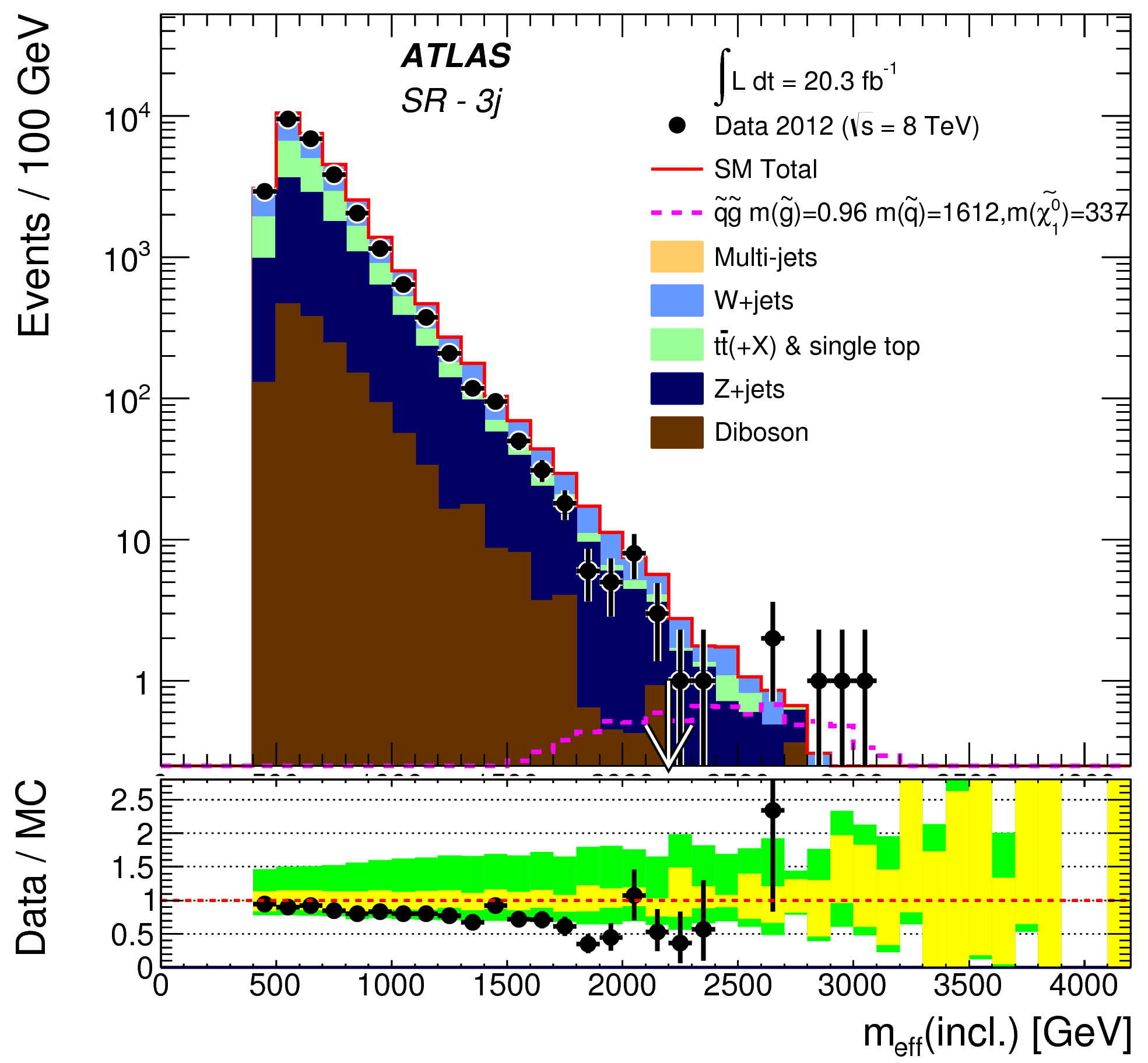}
\end{center}
\caption{$m_{\mathrm{eff}}$ distribution in one signal region of the inclusive search channel with no lepton~\protect\cite{ATLAS_SUSY_0lep_2012}.}
\label{fig:SUSYStrategy2}
\end{figure}

At LHC, TeV-scale squarks and gluinos decay promptly in long decay chains containing mainly quark/gluon jets and the LSP. SUSY events are 
therefore characterized by multiple energetic jets as well as transverse missing energy $E_{T}^{miss}$ originated from the undetected LSP energies. Depending on the 
sparticle mass spectrum between the squarks/gluinos and the LSP, charged lepton(s) and/or photons could also appear in the cascade. Since LHC is an hadronic machine, 
the experimental challenge is to reduce the multijet SM backgrounds by several orders of magnitude. The latter 
is mainly composed of QCD multijets ($\sigma \sim 10^{10}$ pb), $W/Z$+jets ($\sigma\sim 10^{5}$ pb), top ($\sigma\sim 10^{2}$ pb), Dibosons ($\sigma\sim 50$ pb) and 
eventually $ttW$ or $ttZ$ ($\sigma \sim 0.1$ pb). In all cases, the presence of jets and real or fake $E_{T}^{miss}$ could mimic the SUSY signal. 
On top of high initial kinematic cuts, the RPC Strong production analyses are based on very powerful discriminating variables,
which exploit the main characteristics of the SUSY decay chain: the correlation between the scalar sum of the transverse energy of reconstructed objects, $H_T$, and
the module of their vectorial sum, $E_{T}^{miss}$. Combinations of $E_{T}^{miss}$ and $H_T$ like in the effective mass variable, 
$m_{\mathrm{eff}}=H_T+E_{T}^{miss}$~\cite{MEff}, or the missing transverse momentum significance, $E_{T}^{miss}/\sqrt{H_T}$ could provide extra sensitivity. All these 
discriminating variables are consistently used in ATLAS for strong SUSY searches.
They can be linked to some characteristic SUSY parameters like $M_{\mathrm{SUSY}}$, the mass of the highest colored object, $M_{\mathrm{LSP}}$, the LSP mass, 
and their difference $\Delta M$, see Fig.~\ref{fig:SUSYStrategy1}. Typically $m_{\mathrm{eff}}$ will peak at 
$1.8(M_{\mathrm{SUSY}}^2-M_{\mathrm{LSP}}^2)/M_{\mathrm{SUSY}}$~\cite{ATLAS_meff}. For open spectra ($\Delta M> \mathrm{O}(500)$ GeV), this value is well above
the SM background which has no correlation between $E_{T}^{miss}$ and $H_T$, and therefore peaks at lower values, see 
Fig.~\ref{fig:SUSYStrategy2}. However for compressed spectra ($\Delta M<500$ GeV), $m_{\mathrm{eff}}$ loses his separation power and cut 
values should be relaxed. 

\begin{figure}[htbp]
\begin{center}
\includegraphics[height=6cm]{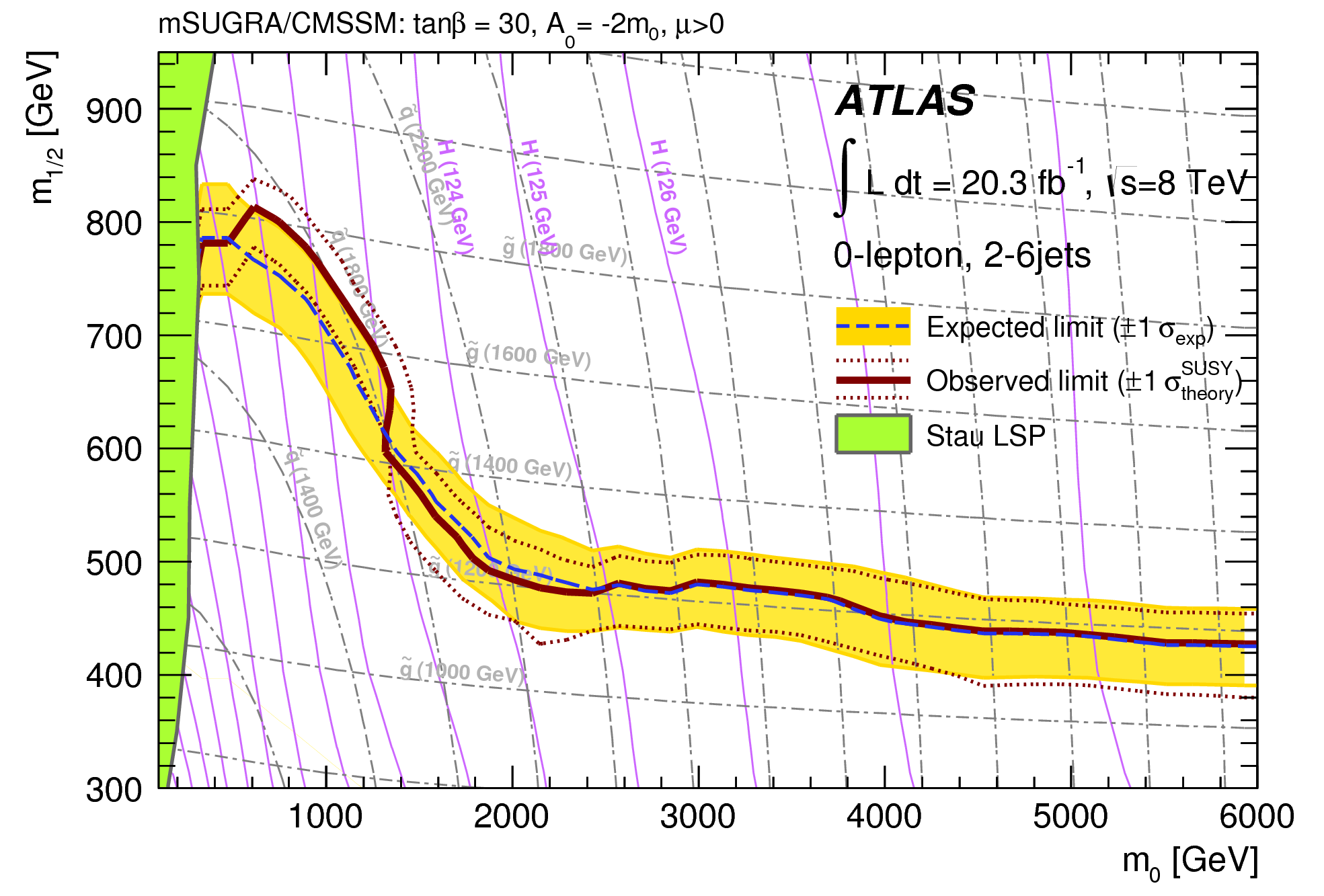}
\end{center}
\caption{SUSY limits from the inclusive search with no lepton in the minimal SUGRA scenario~\protect\cite{ATLAS_SUSY_0lep_2012}.}
\label{fig:SUSYIncl1}
\end{figure}

\begin{figure}[htbp]
\begin{center}
\includegraphics[height=6cm]{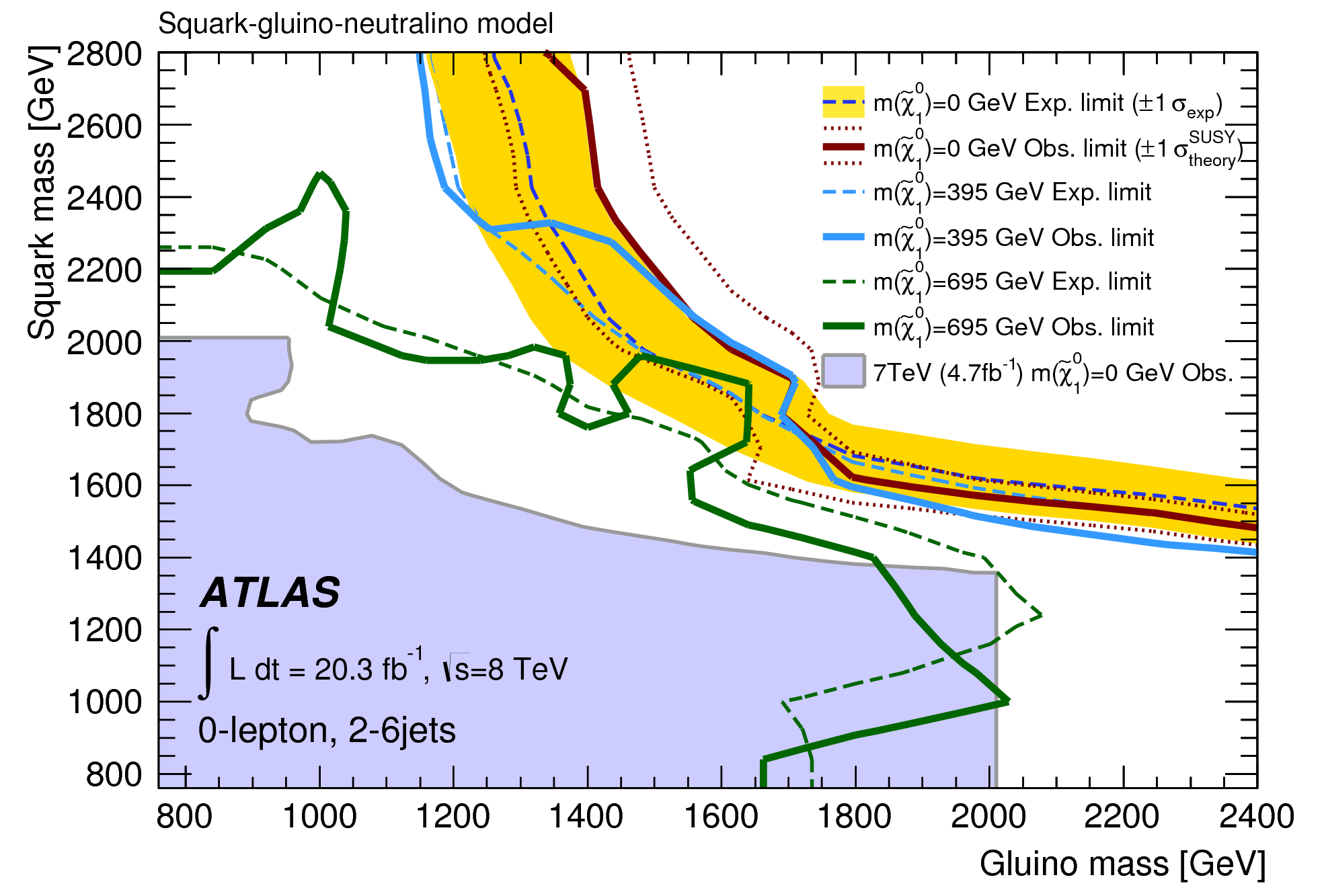}
\end{center}
\caption{Limits on topological models assuming only mass-degenerate first/second generation of squarks, gluinos and the LSP~\protect\cite{ATLAS_SUSY_0lep_2012}.}
\label{fig:SUSYIncl2}
\end{figure}

SUGRA-like scenario can be investigated in the most inclusive way by considering lepton veto analyses. In this case, signatures with 
2 to 6 jets can probe squark-squark (2 jets or more), squark-gluino (3 jets or more) or gluino-gluino (4 jets or more) 
production. Three signal regions (SRs) target the high $M_{\mathrm{SUSY}}$ and high $\Delta M$ by applying tight cuts on 
$m_{\mathrm{eff}}>\mathrm{O}(1)~\mathrm{TeV}$. Seven `medium/loose' SRs cover more compressed spectra, by relaxing the cuts on $m_{\mathrm{eff}}$. 
The absence of excess can be interpreted in a minimal SUGRA (mSUGRA) with 5 parameters (Fig.~\ref{fig:SUSYIncl1}). The two most relevant 
parameters are the universal scalar (fermion) masses at GUT scale called $m_0$ ($m_{1/2}$) proportional to squarks (gluinos) masses at the 
EW scale as shown by the isolines. The other parameters are chosen to accommodate a 125 GeV Higgs mass. The top left part of the plot is 
dominated by squark-squark production best covered by the 4-jets tight SR (2 squark jets + initial and final radiation jets), while the bottom right part, 
is dominated by gluino-gluino, where 6-jets tight SR is best. It is also very interesting to interpret this results 'topologically', i.e. assuming that only few particles 
are accessible in the SUSY mass spectrum. Figure~\ref{fig:SUSYIncl2} show the limits obtained when considering only gluinos, mass-degenerate first/second generation 
squarks and the LSP. Decays of squarks and gluinos are then forced, with 100\% branching ratio, via $\tilde{g} \rightarrow q \bar{q} \tilde{\chi}_1^0$ and 
$\tilde{q} \rightarrow q \tilde{\chi}_1^0$. In this case, mass limits are above 1.4 TeV for gluinos and first/second generation squarks if LSP masses are below 400 GeV. 
For mSUGRA and topological models, squark and gluino with degenerate masses are excluded below 1.7 TeV~\cite{ATLAS_SUSY_0lep_2012}. 

The most 'natural' decay of the gluino is $\tilde{g} \rightarrow t \tilde{t} \rightarrow t t \tilde{\chi}_1^0$, see Fig.~\ref{fig:SUSYGene2}. 
This channel provides 4 tops plus high $E_T^{miss}$ final states but is only poorly covered by the previous analysis. An extra sensitivity can be obtained 
by considering final states not produced by the dominant $t\bar{t} \rightarrow W^+ W^- b \bar{b}$ background, i.e. $i)$ more than 6 jets and 
no lepton~\cite{ATLAS_SUSY_Multijet0l_2012}, $ii)$ two same-sign leptons~\cite{ATLAS_SUSY_2lSS_2012} or $iii)$ 3 $b$-jets~\cite{ATLAS_SUSY_3b_2012} with 
or without a lepton. The best sensitivity is obtained by the latter that can exclude gluino masses up to 1.4 TeV for LSP masses below 500 GeV assuming a 
100\% branching ratio for the decay $\tilde{g} \rightarrow t \tilde{t} \rightarrow t t \tilde{\chi}_1^0$.

All described analyses probe high gluino and squark masses but generally requires open SUSY spectra. For more compressed ones, the $m_{\mathrm{eff}}$ cut could 
be relaxed (medium and loose SRs of the no-lepton analysis) but even there, it is generally not possible to probe LSP masses above 500 GeV. To improve on this, 
asking for one soft electron or $\mu$~\cite{ATLAS_SUSY_1lep_2012} or two same sign leptons is generally better. Experimental challenges 
drastically change: lepton triggers can be exploited and cuts on jet kinematics can be reduced. Lowering cuts on $E_T^{miss}$ and $m_{\mathrm{eff}}$ is possible 
since the multijet QCD background is naturally suppressed by the presence of isolated leptons. Finally other variables exists like the transverse mass $m_T$, 
built from lepton and $E_T^{miss}$, which efficiently reduce $t\bar{t}$ and $W+$jets backgrounds by asking $m_T>m_W$.
These analyses are sensitive to LSP masses of 600 GeV, when the lepton(s) originates from an intermediate sparticle, located between squarks/gluinos and LSP 
like in $\tilde{\chi}_1^{\pm} \rightarrow W^{\pm}(\rightarrow l\nu) \tilde{\chi}_1^0$ or 
$\tilde{t} \rightarrow t \tilde{\chi}_1^0 \rightarrow W^{\pm}(\rightarrow l\nu) b \tilde{\chi}_1^0$. 

GMSB-like scenario are providers of SUGRA-like final states but also offers other experimental possibilities. In these models, LSP is the gravitino and final 
states are driven by the coupling of the Next to Lightest SUSY Particle (NLSP) to the LSP. A natural solution is that the NLSP is the lightest neutralino, 
$\tilde{\chi}_1^0$. In this case, depending on the mixing parameters, $\tilde{\chi}_1^0 \rightarrow \gamma \tilde{G}$, $\tilde{\chi}_1^0 \rightarrow Z^0 \tilde{G}$ and 
$\tilde{\chi}_1^0 \rightarrow h^0 \tilde{G}$ could be opened giving for example $\gamma \gamma$, $\gamma Z^0 (\rightarrow ll)$, $\gamma h^0 (\rightarrow bb)$ final 
states. These final states will be overlaid with jets coming from the gluino/squark cascade. Present limits generally exclude 
gluino masses below 1 TeV and less stringent limits are obtained for squarks~\cite{ATLAS_SUSY_Photon_2011}. However no 8 TeV results are (yet) available.

%-----------------------------------------------------------------------
\subsubsection{Direct searches for third-generation squarks}
\label{sec:SUSY_stop}

Given the limits on the gluino masses, it is conceivable that strong SUSY production could be dominated by the direct production of stop or left sbottom.
Compared to gluino pair-production, less complex final states with an enhanced presence of $b$-jet(s) are then expected. 

\begin{figure}[htbp]
\begin{center}
\includegraphics[height=7cm]{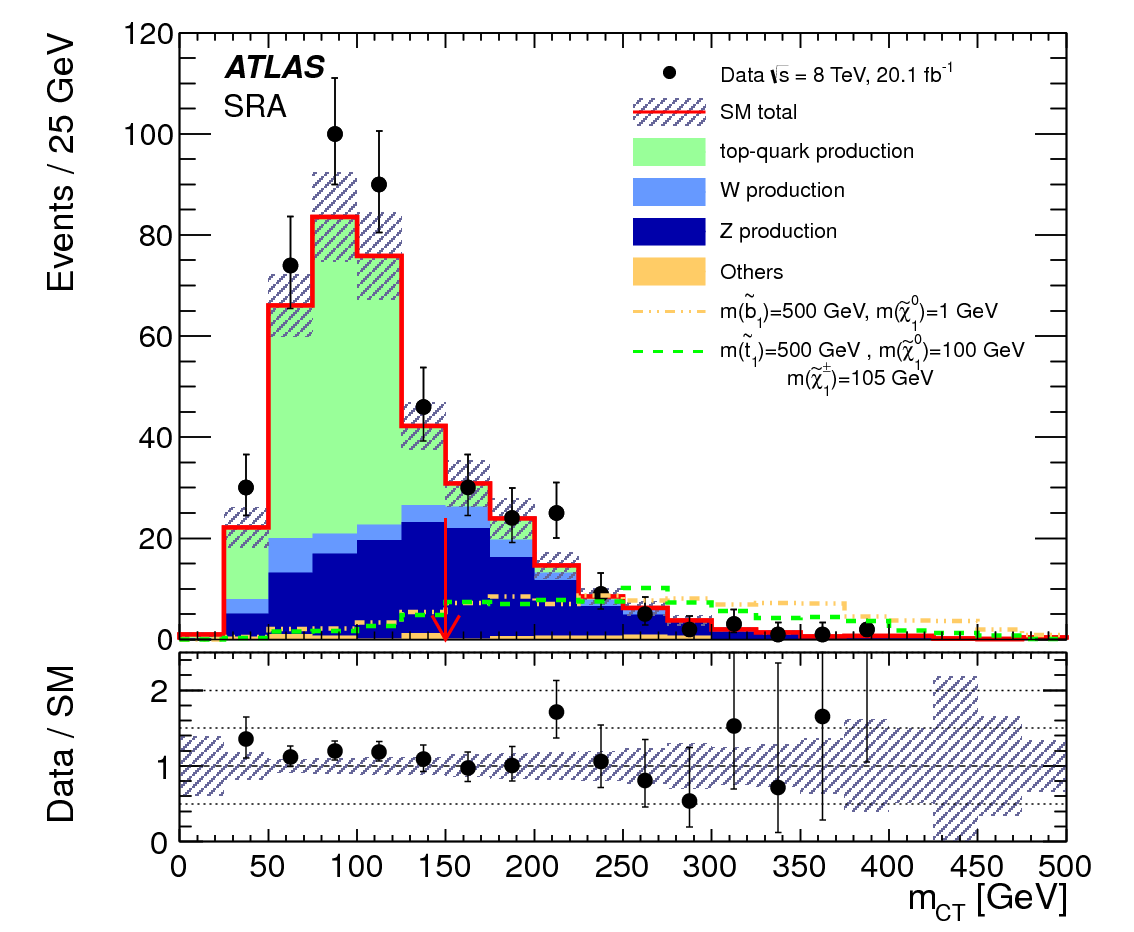}
\end{center}
\caption{Highlights from the direct sbottom searches : $m_{\mathrm{CT}}$ distribution~\protect\cite{ATLAS_Sbottom_2b_2012}.}
\label{fig:SUSYSbottom1}
\end{figure}

\begin{figure}[htbp]
\begin{center}
\includegraphics[height=7cm]{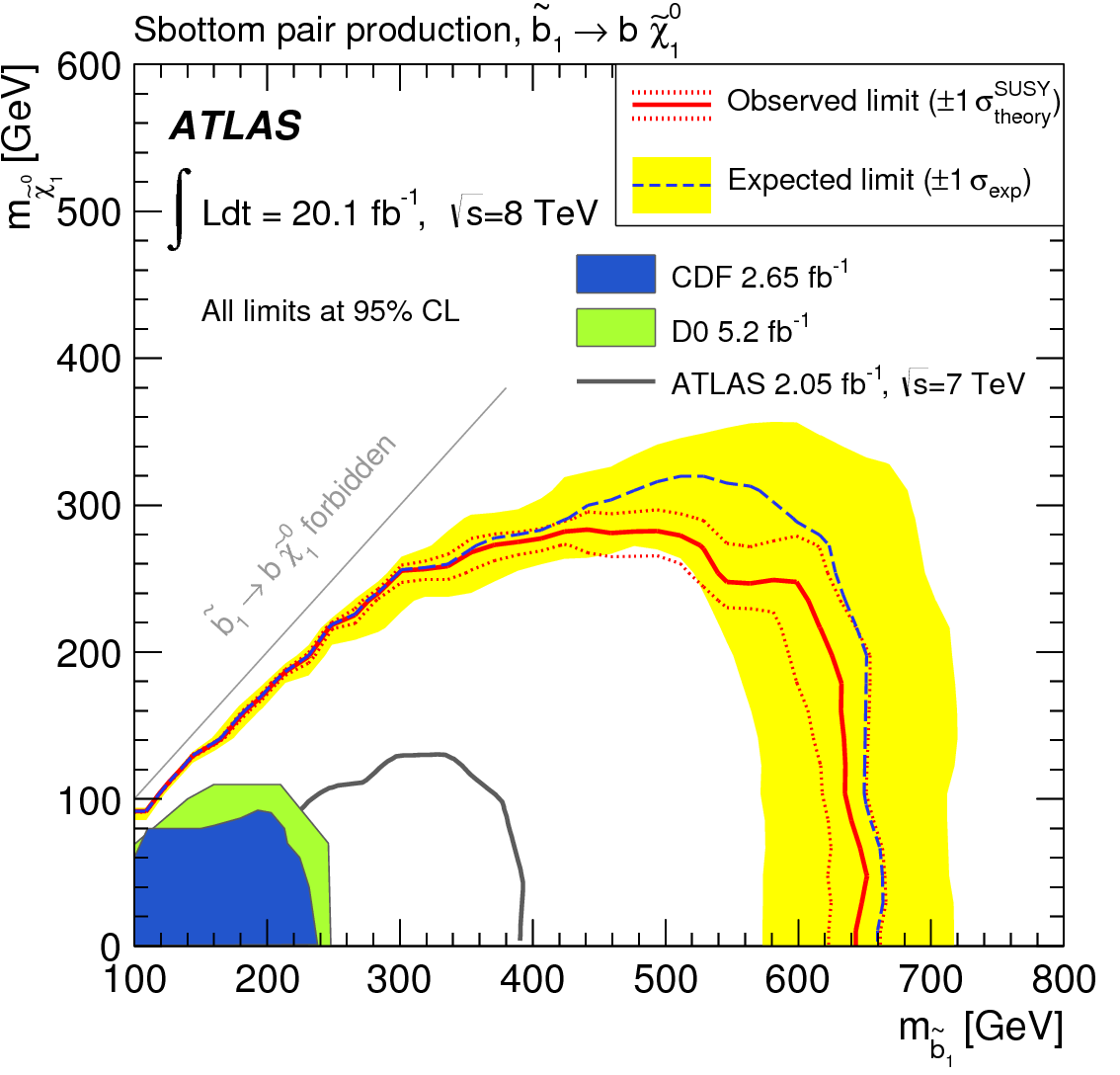}
\end{center}
\caption{Present limits in the sbottom-LSP plane in the case 
$\tilde{b}_L \rightarrow b \tilde{\chi}_1^0$~\protect\cite{ATLAS_Sbottom_2b_2012}.}
\label{fig:SUSYSbottom2}
\end{figure}

The simplest signature is given by the direct left-handed sbottom pair production, where $\tilde{b}_L \rightarrow b \tilde{\chi}_1^0$: exactly two $b$-jets, no lepton and 
high $E_T^{miss}$. To take full advantage of the simple topology, $m_{\mathrm{eff}}$ is replaced by $m_{\mathrm{CT}}$~\cite{ATLAS_SUSY_MCT1,ATLAS_SUSY_MCT2} 
which allows a better signal to background separation, as shown in Fig.~\ref{fig:SUSYSbottom1}. The reason is that two distinct end-points are obtained 
for signal, around $m_{\mathrm{CT}}\sim [M(\tilde{b})^2-M(\tilde{\chi}_1^0)^2]/M(\tilde{b})$, 
and for $t\bar{t}$ background, around $m_{\mathrm{CT}}\sim [M(t)^2-M(W)^2]/M(t)\sim 140$ GeV. Figure~\ref{fig:SUSYSbottom2} shows that sbottom masses below 650 GeV 
are excluded when $M_{\mathrm{LSP}} < 200-300$ GeV~\cite{ATLAS_Sbottom_2b_2012}, getting close to the upper bounds of the natural spectrum. Other sbottom decays 
$\tilde{b} \rightarrow t \tilde{\chi}_1^{\pm}$ and $\tilde{b} \rightarrow b \tilde{\chi}_2^0$, are covered by two same-sign leptons 
and 3 $b$-jets analyses respectively and give generally slightly lower limits.

\begin{figure*}[htbp]
\begin{center}
\includegraphics[height=12cm,width=14cm]{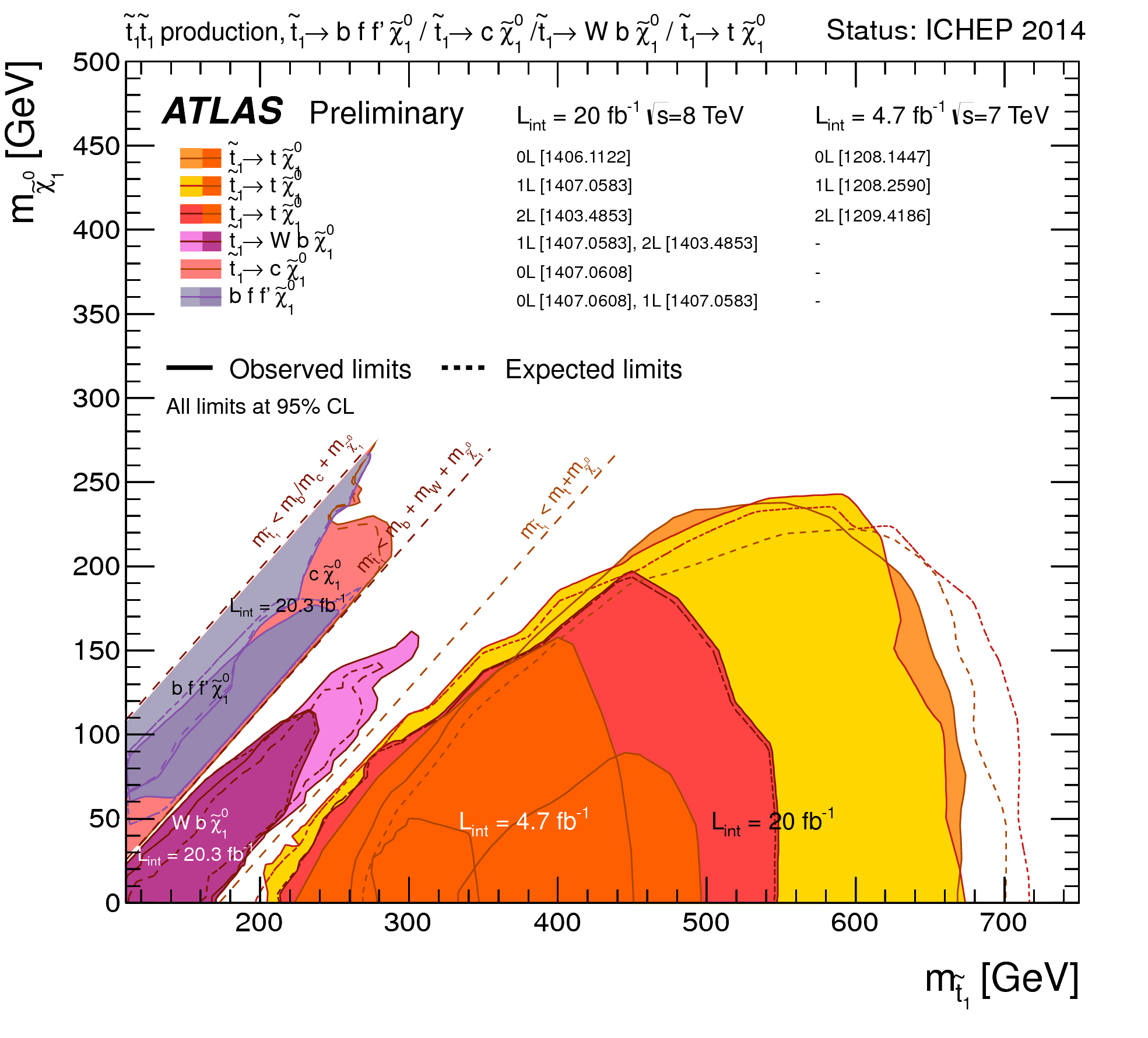}
\end{center}
\caption{Present limits in LSP-stop mass plane.}
\label{fig:SUSYStop}
\end{figure*}

The case of the lightest stop $\tilde{t}_1$, the most pressing issue for the Higgs mass stability at 
high energy, is a bit more complex since the topology is even closer to the SM $t\bar{t}$. Generally the main difference arose from the higher 
expected $E_T^{miss}$ due to the presence of the LSP. The stop decays can be divided in two classes: 
$\tilde{t}_1 \rightarrow t \tilde{\chi}_1^0, b W \tilde{\chi}_1^0, c \tilde{\chi}_1^0$ and 
$\tilde{t}_1 \rightarrow b \tilde{\chi}_1^{\pm} \rightarrow b W^{\pm(*)} \tilde{\chi}_1^0$. The former is best covered by signatures with 0-lepton+6-jets 
including 2$b$-jets~\cite{ATLAS_SUSY_StoptN1_0l} and 1-lepton+4-jets including 1$b$-jet~\cite{ATLAS_SUSY_StoptN1-bC1_1l} for 
$\tilde{t} \rightarrow t \tilde{\chi}_1^0$ decay, 2-leptons+jets for $\tilde{t} \rightarrow b W \tilde{\chi}_1^0$ decay~\cite{ATLAS_SUSY_StopbWN1-bC1_2l} and 
2$c$-jets for $\tilde{t} \rightarrow c \tilde{\chi}_1^0$ decay~\cite{ATLAS_SUSY_StopCharm}. As can be seen from the right part of Fig.~\ref{fig:SUSYStop}, a 
stop mass below 700 GeV is excluded for $M_{\mathrm{LSP}}<100$~GeV, apart from some holes around $m_{\tilde{t}_1} \sim m_t+m_{\tilde{\chi}_1^0}$ because of very 
close topology with $t\bar{t}$. The exclusion weakens for higher LSP masses. The search for 
$\tilde{t}_1 \rightarrow b \tilde{\chi}_1^{\pm}$ depends on the value of the chargino mass. For close-by stop and $\tilde{\chi}_1^{\pm}$, two hard leptons 
are present in the final state and previously mentioned 2-leptons+jets analysis can be reused. If instead $\tilde{\chi}_1^0$ and $\tilde{\chi}_1^{\pm}$ are close-by, 
two hard $b$-jets will be present giving good sensitivity to direct sbottom analysis. Finally if $\tilde{\chi}_1^0$ and $\tilde{\chi}_1^{\pm}$ are far apart a mix of 
1-lepton+4-jets and 2-leptons+jets is best. In the $\tilde{t}_1 \rightarrow b \tilde{\chi}_1^{\pm}$ scenario, the stop could be excluded up to 600~GeV but the 
message is less strong than for the right part of Fig.~\ref{fig:SUSYStop} since it depends on $\tilde{\chi}_1^{\pm}$ masses. 
Nevertheless in this case also stop masses are getting dangerously close to the upper bounds of the natural spectrum.
 
%-----------------------------------------------------------------------
\subsubsection{Direct searches for EWKinos and sleptons}
\label{sec:SUSY_EWK}

Natural SUSY spectrum favors EWKinos close to the EW scale. Therefore they could well be the only sparticles accessible at LHC if colored ones are too heavy 
or decaying through intricate chains. The SUSY electroweak sector is characterized by low cross-sections and discoveries can only happen for lightest mass states 
($\tilde{\chi}_1^0$, $\tilde{\chi}_2^0$ and $\tilde{\chi}_1^{\pm}$) and leptonic final states. 
In the most `natural' scenario, $\tilde{\chi}_1^0$, $\tilde{\chi}_2^0$ and $\tilde{\chi}_1^{\pm}$ are higgsino-like, i.e.
almost mass degenerate. Therefore, for $\tilde{\chi}_1^0$ LSP models, when $\tilde{\chi}_1^{\pm}$ and $\tilde{\chi}_2^0$ decay to $\tilde{\chi}_1^0$, final states are composed of 
low energetic particles, hardly distinguishable from background. As a result, no limits exist on this scenario. Note that for $\tilde{G}$ LSP models, this constraint 
disappears since the mass difference between EWKinos and the $\tilde{G}$ is always significant, generating an interesting phase space, still poorly explored.

\begin{figure}[htbp]
\begin{center}
\includegraphics[height=8cm]{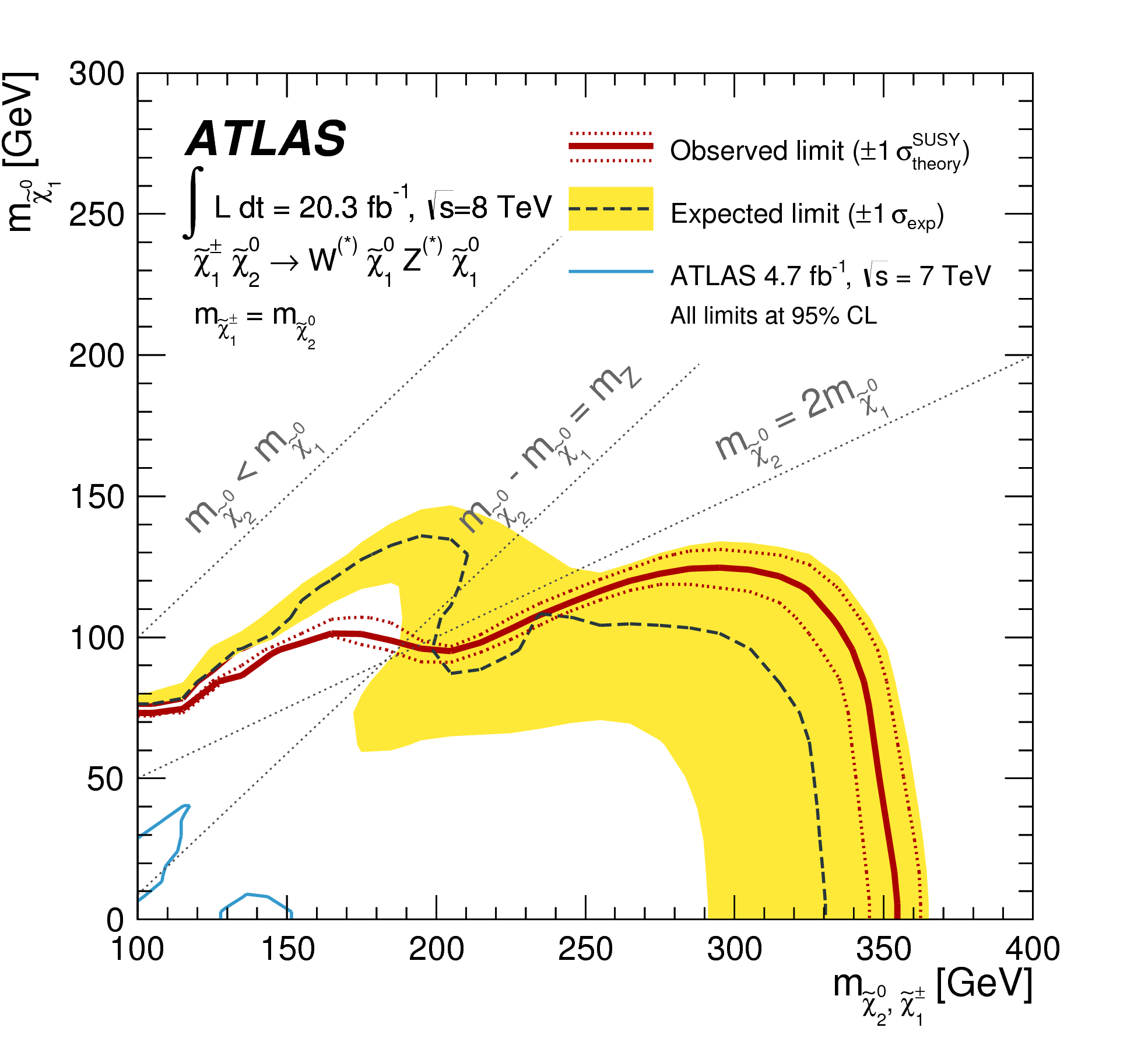}
\end{center}
\caption{Limits for associate EWKinos production as a function of the LSP mass~\protect\cite{ATLAS_SUSY_3lDG}.}
\label{fig:SUSY_EWK_AMSB1}
\end{figure}

Another more favorable scenario consisting in bino-like $\tilde{\chi}_1^0$ and wino-dominated $\tilde{\chi}_2^0$ and $\tilde{\chi}_1^{\pm}$ provides 
a viable solution when $\mu$ is not too high. 
In this case, the mass difference between $\tilde{\chi}_2^0$ and $\tilde{\chi}_1^{\pm}$ and the LSP increases, opening channels like
$\tilde{\chi}_2^0 \rightarrow Z^0 (H^0) \tilde{\chi}_1^0$ and $\tilde{\chi}_1^{\pm}\rightarrow W^{\pm} (H^{\pm}) \tilde{\chi}_1^0$ 
with on-shell $Z$, $H$ and $W$. The highest cross-section is coming from 
$\tilde{\chi}_1^{\pm} \tilde{\chi}_2^0 \rightarrow W (\rightarrow l \nu) Z (\rightarrow l l) \tilde{\chi}_1^0 \tilde{\chi}_1^0$. 
Assuming mass degeneracy between $\tilde{\chi}_1^{\pm}$ and $\tilde{\chi}_2^0$, a 3-lepton+$E_T^{miss}$ analysis excludes 
$M(\tilde{\chi}_1^{\pm},\tilde{\chi}_2^0)$ $<320$ GeV for LSP mass lower than 100 GeV, see Fig.~\ref{fig:SUSY_EWK_AMSB1}~\cite{ATLAS_SUSY_3lDG}. 
Note that breaking the mass degeneracy between $\tilde{\chi}_1^{\pm}$ and $\tilde{\chi}_2^0$ will relax these upper bounds.

Searching for $\tilde{\chi}_1^+ \tilde{\chi}_1^- \rightarrow W^+ (\rightarrow l^+ \nu) W^- (\rightarrow l^- \nu)\tilde{\chi}_1^0 \tilde{\chi}_1^0$ in the 
2-lepton+$E_T^{miss}$ is more challenging because $\sigma(WW)$ $=10 \times \sigma(\tilde{\chi}_1^+\tilde{\chi}_1^-)$ for 100~GeV charginos. 
Vetoing jets and using similar discriminant variables as for direct sbottom searches, the $WW$ 
background can be reduced sufficiently to exclude few SUSY models~\cite{ATLAS_SUSY_2lDG}. An interesting by-product of these searches 
is the possibility to exclude sleptons (selectron and smuon) at higher masses than at LEP, since both signal have the very same 2-lepton+$E_T^{miss}$ final state 
($\tilde{l}^+ \tilde{l}^- \rightarrow l^+ l^-  \tilde{\chi}_1^0 \tilde{\chi}_1^0$).

Finally $\tilde{\chi}_1^+ \tilde{\chi}_1^0 \rightarrow W^+ (\rightarrow l^+ \nu) \tilde{\chi}_1^0$ is not yet explored due to the very low cross-section 
and overwhelming inclusive $W$ cross-section. For the same reasons, the $\tilde{\chi}_1^0 \tilde{\chi}_1^0$ production cannot be searched for, even in dedicated 
monojet analyses presented in Section~\ref{sec:ADD}. In conclusion, EWKino searches provide presently much weaker constraints on the natural SUSY scenario.

%-----------------------------------------------------------------------
\subsubsection{Status of R-Parity Conserved SUSY after LHC run I}
\label{sec:SUSY_RPC}

LHC have probed the uncharted heart of natural weak scale SUSY spectrum by direct searches. The limits are especially strong for gluinos and $3^{\mathrm{rd}}$ 
generation squark in open spectra. Masses below 1~TeV and 500-700~GeV are excluded, respectively. These constraints set plain vanilla MSSM on the grill and pushed it 
in corners of parameter space harder to access experimentally: compressed spectra and intricate decay chain (for strong SUSY), low cross-section processes 
(especially in the EW sector). To be more quantitative, a full scan of the most relevant 19-20 MSSM parameters (some assumptions are made on the 
other 105-19 parameters) was performed and the models surviving the LHC results examined. This study confirmed that spectrum evading the limits are 
typically the ones containing light stop and bottom with complex decay patterns~\cite{pMSSM1,pMSSM2}.

%-----------------------------------------------------------------------
\subsubsection{Escape routes: Long-Lived particles, R-Parity Violation and others}
\label{sec:SUSY_Escape}

Beside the reasons given above, three possible escape routes could explain the null results in RPC SUSY searches. First, some particles of the SUSY spectrum can be 
metastable, i.e. they have non prompt decays within the inner detector giving non pointing $\gamma$ or $Z$, displaced vertices or disappearing tracks or even decay 
after the detector. Second, R-parity is violated i.e. one/several of the 48 Yukawa couplings in the superpotential~\cite{RPV} is/are non zero. Note that having 
all RPV parameters different from 0 is not possible because of the limits from proton lifetime. This generally gives striking signatures with lepton flavor 
violation ($\lambda_{ijk}\neq 0$, $\lambda'_{ijk}\neq 0$, $i,j,k=1,2,3$) or baryon number violation ($\lambda''_{ijk}\neq 0$). The third possibility is a more complicated SUSY model, 
beyond the MSSM, relaxing the experimental constraints: for example a new singlet can be added, enlarging the EW sector with 2 more Higgs bosons and one more 
neutralino~\cite{NMSSM}, or even SUSY could have no role in the hierarchy problem and only the LSP could be present at the TeV-scale, vestige of a very high mass 
scale spectrum~\cite{UnnaturalSUSY}. As the phase space is huge and less well-defined than in the RPC case, only some illustrative examples are discussed below.

\begin{figure}[htbp]
\begin{center}
\includegraphics[height=8cm]{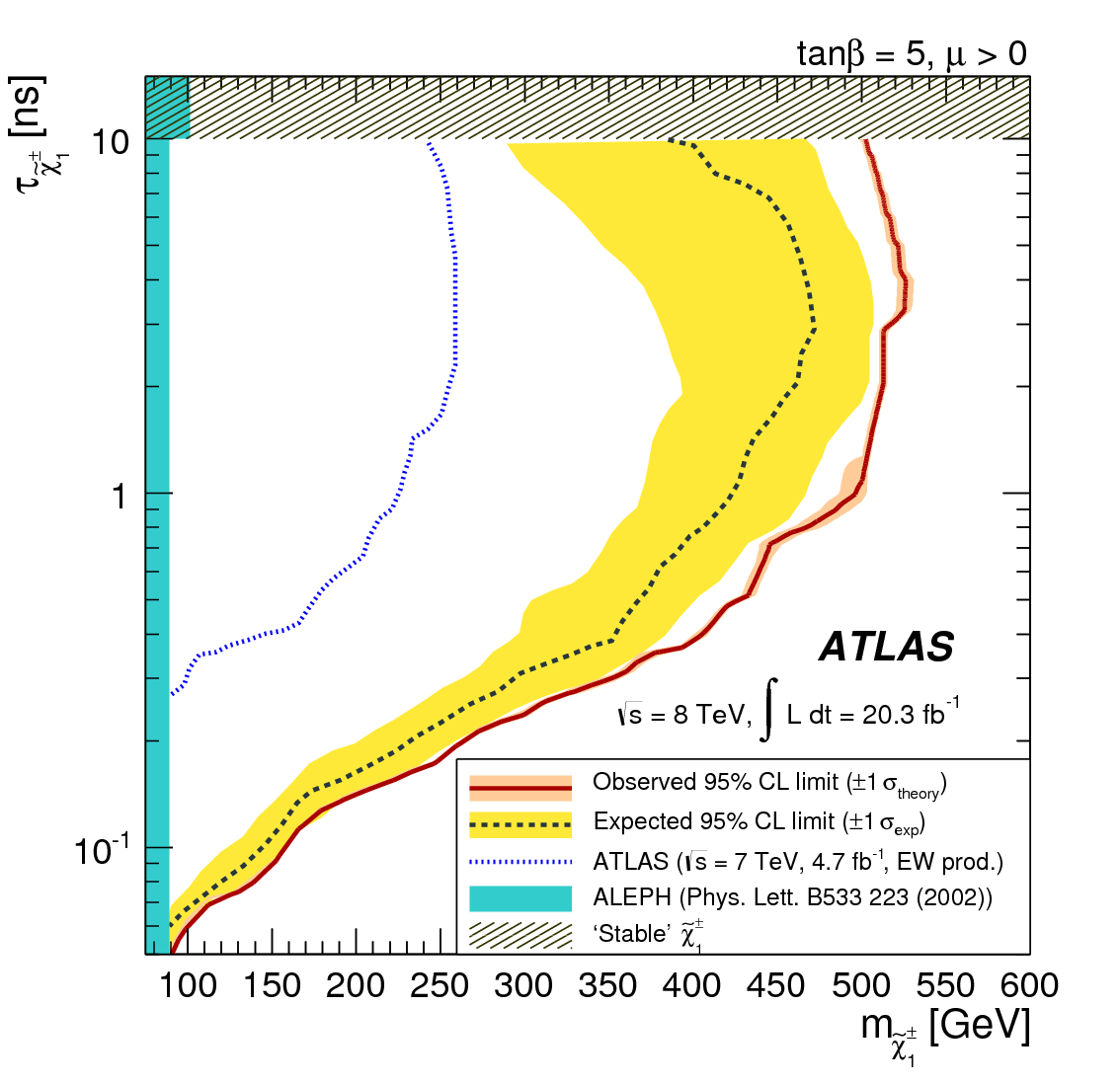}
\end{center}
\caption{Limits in the lifetime-mass plane for a metastable chargino~\protect\cite{ATLAS_SUSY_DC1}.}
\label{fig:SUSY_EWK_AMSB2}
\end{figure}

As no metastable particles are present in the Standard Model their searches are generally background free. Their discovery in-turn requires a deep understanding of the 
detector performance, which represent the only background. In SUSY, non-prompt particle decay can be caused by $i)$ 
very weak R-Parity violation, i.e. one of the Yukawa coupling $\lambda, \lambda'$ or $\lambda'' \leq \mathrm{O}(10^{-5})$, $ii)$
very low mass difference between a SUSY particle and the LSP in RPC model or $iii)$ very weak coupling to the gravitino in GMSB models. AMSB provides a well 
motivated case for $ii)$ where $\tilde{\chi}_1^{\pm}$ and $\tilde{\chi}_1^0$ are almost degenerate and $M(\tilde{\chi}_1^{\pm})-M(\tilde{\chi}_1^0) \geq 140$~MeV.
The chargino is therefore metastable and decays after few tens of centimeters to undetectable particles, a soft pion and the LSP. This will cause the chargino 
track to `disappear'. When produced directly ($\tilde{\chi}_1^+\tilde{\chi}_1^-$, $\tilde{\chi}_1^{\pm}\tilde{\chi}_1^0$) with an additional jet from initial 
state radiation to trigger the event, one (or two) tracks may have no/few associated hits in the outer region of the tracking system. The continuous 
tracking of the outer part of the ATLAS inner detector, the straw tube transition radiation (TRT), gives sensitivity to this signature 
and removes the background. With the additional requirement 
of an high energetic isolated track, regions beyond the LEP limits can be excluded in the lifetime-mass plane of the chargino, 
as shown in Fig.~\ref{fig:SUSY_EWK_AMSB2}~\cite{ATLAS_SUSY_DC1}. Although originally motivated by AMSB, this result is largely model independent and is 
also predicted by unnatural SUSY~\cite{UnnaturalSUSY}.

If gluino and LSP are almost mass degenerate, its lifetime could be long enough for him to hadronize in $R$-hadrons ($\tilde{g}q\bar{q}$, $\tilde{g}qqq$) 
or $R$-gluino balls ($\tilde{g}g$). A fraction of these slow moving particles may come to rest within the detector volume and only decay later as 
$\tilde{g} \rightarrow q\bar{q} \tilde{\chi}_1^0, g \tilde{\chi}_1^0$. If this happens in the calorimeter, the signature will be high energetic jet(s) in absence of 
collisions. In that case, the background comes from the calorimeter noise burst, cosmic ray with high energy deposit or beam halo -- the leading background. 
Gluinos below 850 GeV are excluded for a gluino lifetime between 10 $\mu s$ and 15 minutes~\cite{ATLAS_StoppedGluinos}.

Sizable R-Parity violation in one of the Yukawa coupling ($\sim 10^{-4,-2}$) can easily give four or more leptons because of the 
LSP decay (e.g. $\lambda_{121}\neq 0$) or 2$\times$3-jet resonances ($\lambda''_{ijk}\neq 0$). In the former case, assuming 
$\tilde{g} \tilde{g} \rightarrow qq \tilde{\chi}_1^0 (\rightarrow l^+l^- \nu) qq \tilde{\chi}_1^0 (\rightarrow l^+l^- \nu)$ allows to exclude gluino below 1.4 TeV
~\cite{ATLAS_SUSY_RPV_4lDG} while in the latter case gluino just below 1 TeV are excluded by asking 6-jets to be reconstructed with a minimum energy as 
expected from $\tilde{g} \tilde{g} \rightarrow \tilde{q}(\rightarrow qq)q \tilde{q}(\rightarrow qq)q$~\cite{ATLAS_SUSY_RPV_6j}.

In all these models it is generally true that the gluino mass is excluded for masses below 1 TeV.

%&&&&&&&&&&&&&&&&&&&&&&&&&&&&&&&&&
\subsection{Searches for other natural theories at LHC}
\label{sec:BSM_NonSUSY}
%&&&&&&&&&&&&&&&&&&&&&&&&&&&&&&&&&

As discussed in Section~\ref{sec:Higgs_NP}, alternatives to SUSY exist to solve the hierarchy problem. They generally have distinct 
features compared to SUSY signatures: higher cross-section, moderate or null $E_{T}^{miss}$, high mass resonances decaying to very energetic calorimeter objects 
(electrons, photon and jets) or boosted top, $W$ or $Z$. In the latter case, very collimated objects are produced and reconstructed into one single ``fat" jet. 
Using the high granular ATLAS and CMS detectors, dedicated algorithms were developed to look for substructure in very high energetic jets and to separate the initial 
objects. These methods greatly improve the reconstructed top, $W$ or $Z$ mass resolution, increasing the sensitivity to new physics.

%-----------------------------------------------------------------------
\subsubsection{Large Extra Dimensions}
\label{sec:ADD}

The most striking possibility is that gravity is strong close to the EW scale. Assuming its lines of force propagate in $4+\delta$ large flat extra 
spatial dimensions (the "bulk"), gravity will be ``artificially" weak in our four dimensional rigid brane, where SM particles are confined. In this model, called ADD~\cite{ADD},
the hierarchy problem needs to be rewritten. A $\delta$-dimension fundamental Planck mass, $M_D$, can be computed as a function of the 
compactification radius $R$ of the extra dimensions on a $\delta$-dimensional torus or a sphere as:
\begin{equation}
M_D = \left[\frac{M_{Pl}^2}{R^{\delta}}\right]^{-\frac{1}{2+d}}
\end{equation}
If $M_D \sim 1$ TeV, the hierarchy problem is solved and as $\hbar c=2\times 10^{-14}$ GeV.cm,
\begin{equation}
R=2\times 10^{-17+32/\delta}~\mathrm{cm}
\end{equation}
Only $\delta=1$ is excluded experimentally. In the bulk, gravitational interaction are mediated by massless graviton and 
Kaluza-Klein (KK) graviton towers $G^{(k)}$ are predicted in the 4D brane with masses:
\begin{equation}
m_k^2=m_0^2+k^2/R^2, k=0,1,2,3,4, ...
\end{equation}
For large $R$, the KK states are almost continuous which compensate the small graviton coupling ($\sim 1/M_{Pl}$). Three of the most spectacular 
signatures expected at LHC are now discussed.

\begin{figure}[htbp]
\begin{center}
\includegraphics[height=8cm]{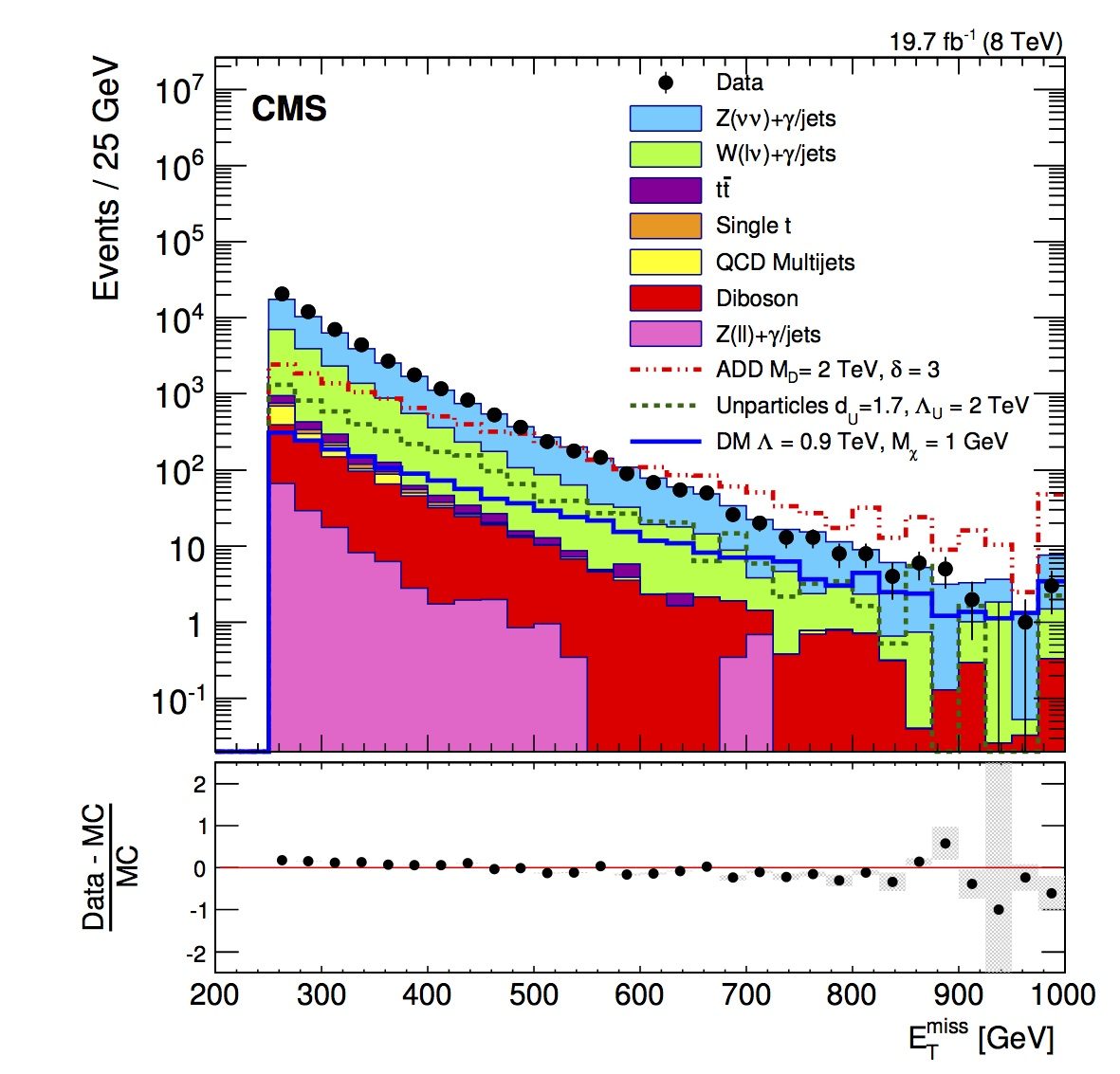}
\end{center}
\caption{ADD model searches at LHC: $E_{T}^{miss}$ distribution in Monojet analysis~\protect\cite{CMS_MonoJet}.}
\label{fig:ED_ADD1}
\end{figure}

\begin{figure}[htbp]
\begin{center}
\includegraphics[height=7cm]{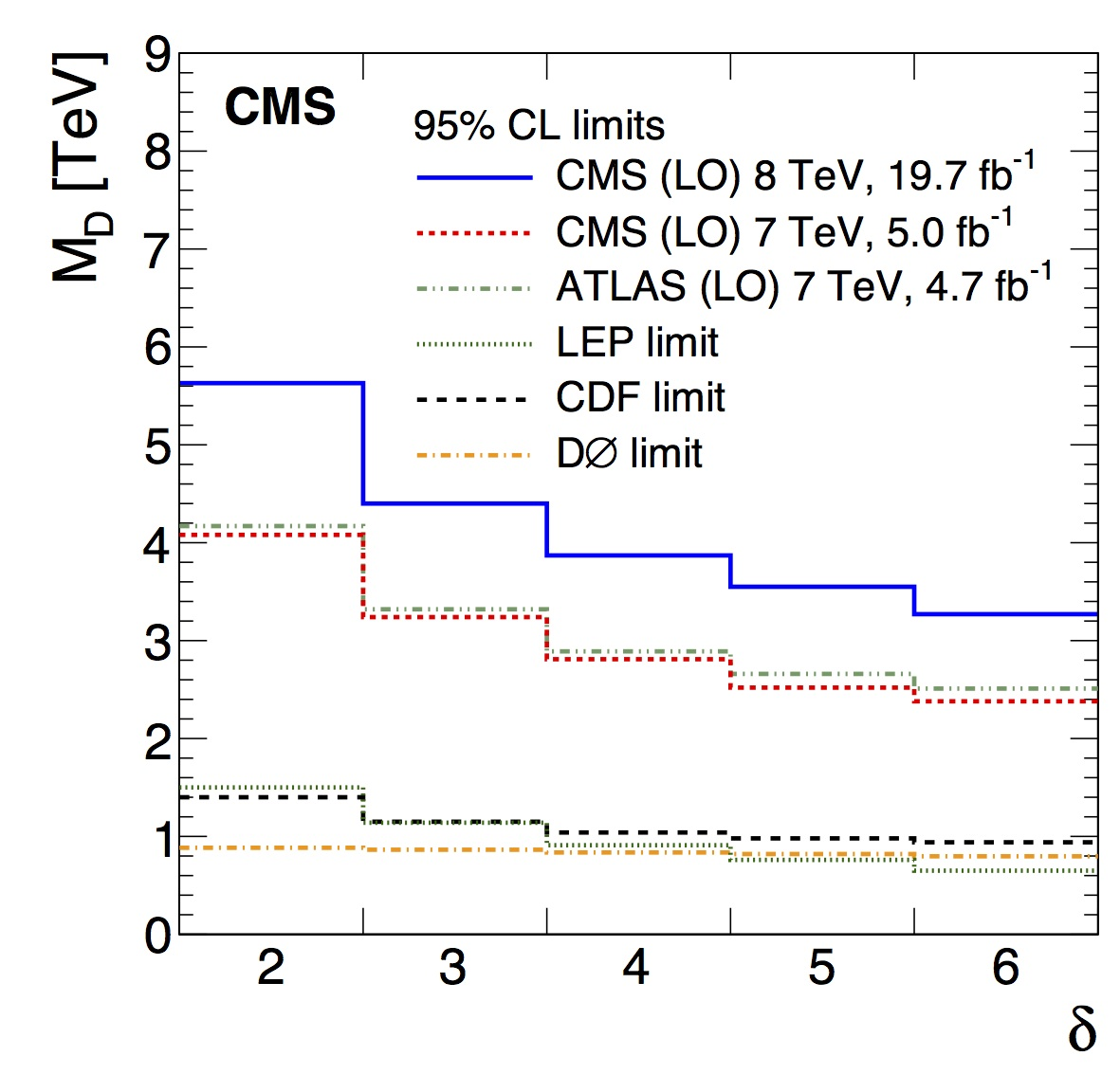}
\end{center}
\caption{Exclusion of ADD Models with Monojet analysis for different $\delta$-dimension~\protect\cite{CMS_MonoJet}.}
\label{fig:ED_ADD2}
\end{figure}

First, the direct production of KK gravitons via the processes $q\bar{q}\rightarrow gG$, $qg\rightarrow qG$, $gg\rightarrow gG$ could provide a monojet signature 
as the graviton escape detection~\cite{Monojet_ADD}. In this scenario a larger tail is expected in the $E_{T}^{miss}$ distribution, see Fig.~\ref{fig:ED_ADD1}, compare 
to the dominant background $Z\rightarrow \nu\nu$ plus one jet emitted in the initial state. The search is limited by the statistics in the 
$Z\rightarrow \nu\nu$ control region where two leptons are required on top of the jet and $E_{T}^{miss}$ kinematic cuts. Nonetheless a huge range of models can 
be excluded with respect to LEP and Tevatron as shown in Fig.~\ref{fig:ED_ADD2}~\cite{CMS_MonoJet}.

\begin{figure}[htbp]
\begin{center}
\includegraphics[height=7cm]{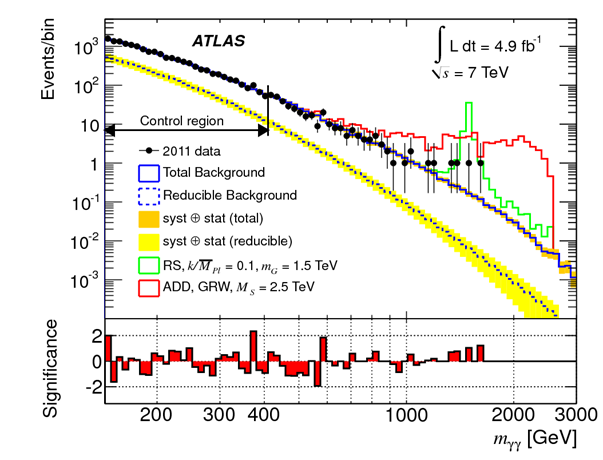}
\end{center}
\caption{ADD model searches at LHC: diphoton resonance~\protect\cite{ATLAS_GammaGamma}.}
\label{fig:ED_ADD3}
\end{figure}

Another interesting signature exists when $s$-channel KK gravitons exchange takes place and KK gravitons decay to dibosons, dileptons and/or dijets, causing large invariant 
masses. One of the most promising signature is the search for diphoton resonance since it is possible to reduce the $\gamma$-jet and jet-jet background below the 
irreducible $\gamma \gamma$ background from SM (Section~\ref{sec:Higgs_Disco}). The ADD signal will appear as a wide resonance over the background, see 
Fig.~\ref{fig:ED_ADD3}~\cite{ATLAS_GammaGamma}. This channel provides similar limits as the monojet analyses for $\delta>2$.

\begin{figure}[htbp]
\begin{center}
\includegraphics[height=9cm]{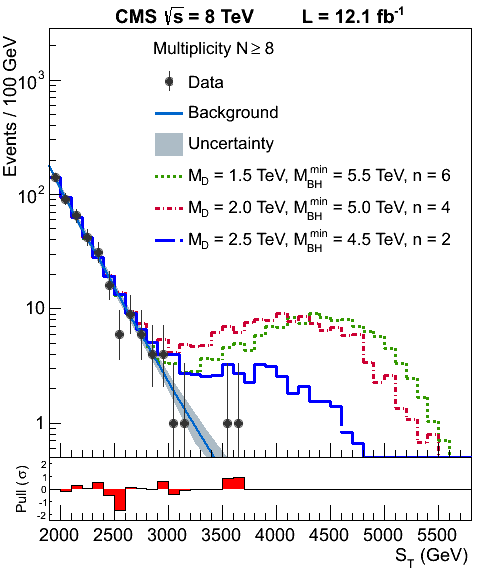}
\end{center}
\caption{ADD model searches at LHC: total transverse energy for $N\ge 8$ objects ($e$, $\mu$, $\gamma$, jets). $S_T$ means $H_T$ here and $n=\delta$~\protect\cite{ADD_ST}.}
\label{fig:ED_ADD4}
\end{figure} 

Finally since $\sqrt{s}>M_D \sim 1$ TeV, gravity is enhanced in the 4+$\delta$ space and microscopic black holes could be produced at LHC. They will then evaporate 
through Hawkings radiation in a high multiplicity ($N$) of particles, see Fig.~\ref{fig:ED_ADD4}. Here the background is dominated by QCD and 
estimated assuming a common shape for $H_T$ regardless of $N$. Large uncertainties exist on these Black Hole (BH) production models due to our ignorance 
of quantum gravity. Assuming (semi-)classical approximation are valid for $M_{BH}>M_D$, quantum black holes with masses below 4.3-6.2 TeV 
are excluded~\cite{ADD_ST}. Other signatures can be used like 2 same sign muons~\cite{ADD_2SSMuon} or even lepton+jets signature~\cite{ADD_1ljets}.

%-----------------------------------------------------------------------
\subsubsection{Warped Extra Dimensions}
\label{sec:WarpedRS}

The hierarchy problem can also be solved by considering only one extremely small new compact dimension with a warped geometry of curvature $k\approx M_{Pl}$ where 
only gravity propagates~\cite{RS}. This set-up, called minimal Randall-Sundrum (RS), is composed of a 5 dimensional bulk with one compactified dimension, 
and two 4D branes, called SM and gravity branes. In these conditions the Planck scale is red-shifted for SM brane observers and becomes 
$M_D=M_{Pl} e^{-k\pi R}$. For $kR \sim 12$, i.e. $R=10^{-32}$~cm, $M_D \sim 1$TeV which solve the hierarchy problem. 

\begin{figure}[htbp]
\begin{center}
\includegraphics[height=6cm]{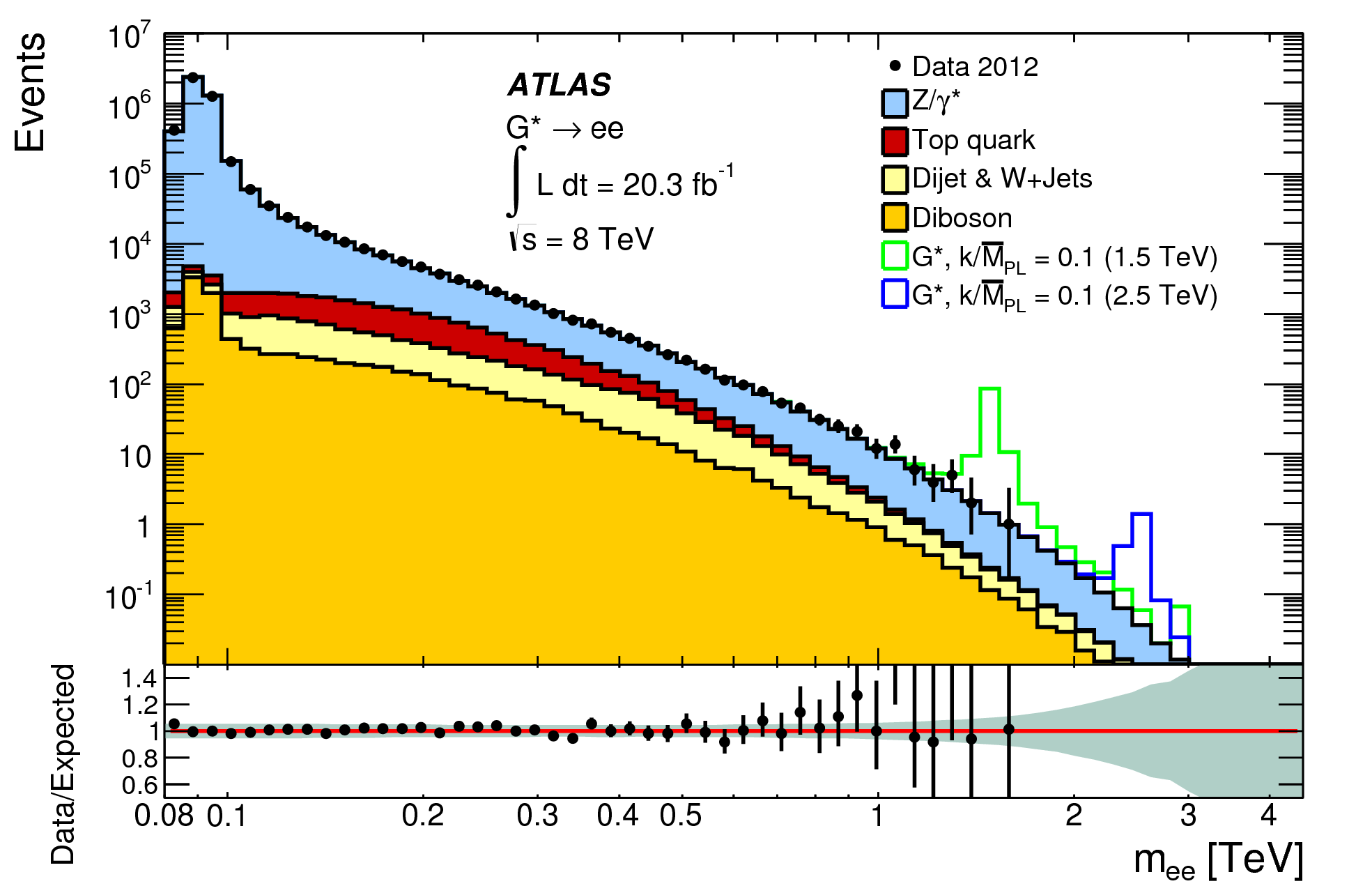}
\end{center}
\caption{Warped extra dimension searches at LHC: dielectron resonance~\protect\cite{Zprime}.}
\label{fig:ED_Warped1}
\end{figure}

Experimental consequences are very different from the ADD case: KK graviton masses are not regularly spaced but given by $m_n=x_n k e^{-k\pi R}$ 
where $x_n$ are the roots of Bessel functions. Only the first excitation, $G^{(1)}$, with a narrow width $k/M_{Pl}<1$, is generally accessible at LHC. 
Its coupling to SM particles is proportional to $1/(M_{Pl} e^{-k\pi R})$ and therefore much stronger than for ADD model. 
As a consequence the main experimental evidence is a narrow peak in the diboson (Fig.~\ref{fig:ED_ADD3}) or dilepton 
(Fig.~\ref{fig:ED_Warped1}) invariant mass. For $k/M_{Pl}=0.02$, $G^{(1)}$ masses below 2.7~TeV are excluded~\cite{Zprime}.

\begin{figure}[htbp]
\begin{center}
\includegraphics[height=7cm]{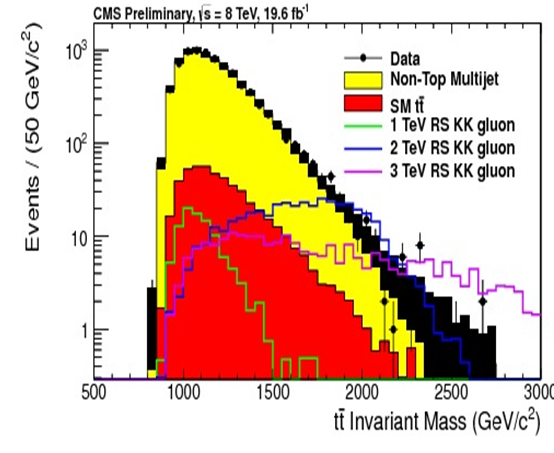}
\end{center}
\caption{Warped extra dimension searches at LHC: $t\bar{t}$ resonance~\protect\cite{RS_ttbar}.}
\label{fig:ED_Warped2}
\end{figure}

Since solving the hierarchy problem requires only the Higgs to be close to the SM brane, the minimal RS can be modified by allowing 
SM fields to propagate also in the bulk~\cite{RS-bulk}. This has the extra advantage to explain the SM Yukawa coupling hierarchies by 
the position of the SM fields in the bulk. All SM fields create KK towers which are 
constrained to have below than 2-3 TeV masses for the first excitation~\cite{RS_Limit}. A particularly interesting search comes from 
the KK gluon ($g_{KK}$) decaying to $t\bar{t}$ which provides an enhancement at high mass of $t\bar{t}$ invariant mass spectrum, as shown 
in Fig.~\ref{fig:ED_Warped2} for the all hadronic channel~\cite{RS_ttbar}. Combining all sensitive $t\bar{t}$ decay channels, $g_{KK}$ masses are 
excluded below 2.5 TeV~\cite{RS_ttbarPubli} getting close to the upper part of the allowed region.
In these RS-bulk models the cross-section of $gg \rightarrow G^{(1)} \rightarrow WW (ZZ)$ is driving the $G^{(1)}$ hunt motivating a search for
$ZZ$ or $WW$ resonances~\cite{RS_ZZ,RS_WW}. However, presently, no mass limit beyond 500 GeV can be put on $G^{(1)}$ when $0.04<k/M_{Pl}<0.1$.

%-----------------------------------------------------------------------
\subsubsection{Composite Higgs models}
\label{sec:CH}

All experimental results indicate today that the new discovered particle with a mass of 125 GeV is the SM Higgs boson. However it is not excluded that 
the Higgs boson is a composite particle. This would then be interpreted as the 
first manifestation of a new strong sector that should appear at a scale $f\sim \mathrm{O}(\mathrm{TeV})$. Note that the Higgs couplings will then be 
modified by $\sim \mathrm{O}(v^2/f^2)$, i.e. remain quite close to the SM values and compatible with the present measurements.
If the Higgs is composite, it could play a similar role as the neutral pion for the strong force, a pseudo Goldstone boson of a spontaneously broken new global 
symmetry. This could explain also why it is much lighter than the other (unobserved) resonances.

Many models have been constructed upon this generic idea, from technicolor to little Higgs models. As of today, composite Higgs models~\cite{CH} are probably the less 
constrained ones. These models also present the advantage to be related by holography to weakly-coupled models containing a warped extra 
dimension, presented just before. The advantage is that perturbative computations can be performed in these extra dimension models and later be used 
to derive the properties of the expected composite states. 

\begin{figure}[htbp]
\begin{center}
\includegraphics[height=7cm]{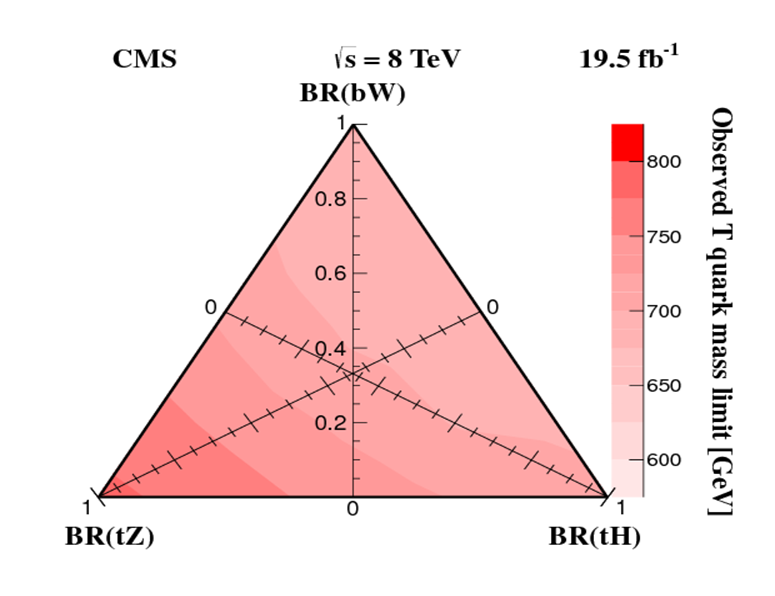}
\end{center}
\caption{Status of $T_{2/3}$ search~\protect\cite{CH_Top23}.}
\label{fig:Other1}
\end{figure}

In composite Higgs models, the Higgs mass is typically around 0.2 TeV or higher but 125 GeV could still be accommodate. The hierarchy problem is solved by 
the finite size of the Higgs, which screens the contributions to its mass from O(TeV) new particles, a similar mechanism as for SUSY. 
These new particles are vector-like top partners and should be light (around 0.7 TeV), or $Z'$ and $W'$ which should be in the 1-3 TeV mass range. As in SUSY models, 
the discovery of these top partners ($T$) is the most pressing issue. To evade the EW precision fit constraint, $T$ could have the form of 
an electroweak singlet of charge 2/3 and searched for via direct production and the subsequent decay to $tW$, $bW$, $tZ$, $tH$. These decays 
generate multi-$W$, i.e 1, 2 and/or 3-lepton+jets final states. When all these channels are combined, it is possible to exclude 
$T_{2/3}$ with masses below 687 GeV as shown in Fig.~\ref{fig:Other1}~\cite{CH_Top23}. Another possibility is $T_{5/3} \rightarrow tW$ and similar 
limits are obtained with a 2 lepton same sign analysis~\cite{CH_Top53}.

\begin{figure}[htbp]
\begin{center}
\includegraphics[height=7cm]{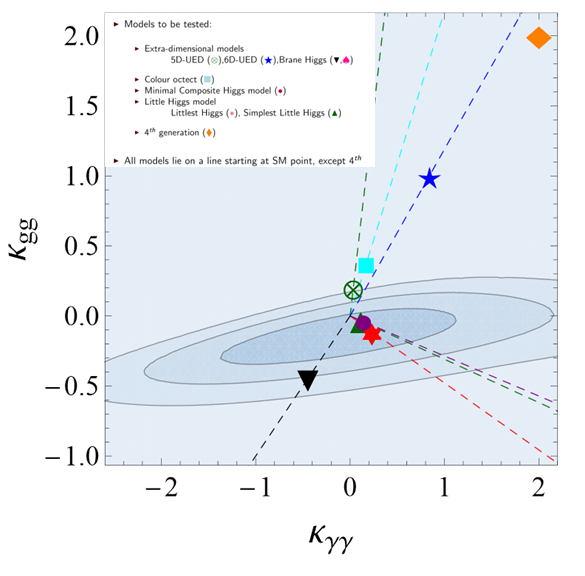}
\end{center}
\caption{Illustration of the $4^{\mathrm{th}}$ generation quark model (diamond on the top right) exclusion by Higgs 
couplings measurements (shaded area)~\protect\cite{4thGene}.}
\label{fig:Other2}
\end{figure}

%-----------------------------------------------------------------------
\subsubsection{Preliminary conclusions on searches for Natural theories}
\label{sec:ConcBSM}

Since 40 years many BSM theories were developed to solve the hierarchy problem with new physics strongly or weakly interacting with the Higgs. This is today the most 
outstanding problem of the SM, getting even stronger with the discovery of a SM Higgs-like particle. Unexpectedly, no sign of new physics have been 
observed with the LHC run I data and most of these BSM theories are now 
seriously cornered. The new predicted particles are now generally excluded close or above the 1 TeV scale. LHC seems to disfavor a 
'natural' scenario, even if all results have not been obtained and some holes still exist in analyses. The complete analysis of the next LHC run at 13-14 
TeV will enable to make a more definitive statement.

%&&&&&&&&&&&&&&&&&&&&&&&&&&&&&&&&&
\subsection{Other Beyond Standard Model searches at LHC}
\label{sec:OthersBSM_noHierarchyPb}
%&&&&&&&&&&&&&&&&&&&&&&&&&&&&&&&&&

As mentioned in Section~\ref{sec:Higgs_NP} there are many other possible SM extensions that do not solve the hierarchy problem but addresses other 
conceptual problems. I just highlight here two very important searches. The first one is probing the quark compositeness by looking for a resonance 
in a dijet invariant mass $m_{jj}$ spectrum~\cite{DiJets} and measuring the relative proportion of central jets per $m_{jj}$ bins. A bump in 
the $m_{jj}$ distribution or an increase of central jets at high $m_{jj}$ could reveal a new substructure~\cite{DiJets_RutherFord}, like the 
gold-foil Rutherford experiment revealed the atomic nucleus. In both cases, null results are obtained and excited quarks below $\sim$ 4 TeV are excluded. 
The second search for the presence of a fourth generation of quarks which can be directly excluded by the Higgs coupling measurements. As discussed 
in Section~\ref{sec:Higgs_Disco}, the Higgs production and decay in $H\rightarrow \gamma \gamma$ channel occurs almost entirely by triangle heavy fermion 
loops. Therefore an enhancement is expected, in both cases, in presence of 4$^{\mathrm{th}}$ generation quarks. As shown in Fig.~\ref{fig:Other2}, these 
models are already excluded~\cite{4thGene}. Results not mentionned here can be can be found on the experiment websites~\cite{CMS_EXO_Res,ATLAS_EXO_Res}.

Since this lecture concentrates on ATLAS and CMS results, it worth to remind that precision measurements and rare decay searches are also very sensitive to
the presence of new physics far below the TeV scale: not as a direct evidence but as deviations from SM expectations that could be explained by ``virtual" 
effects including new physics. Most powerful probes are provided by the proton decay, lepton flavor violation, FCNC in the quark sector, 
electric dipole momentum~\cite{IntensityFrontier,LowEnergy}.

%&&&&&&&&&&&&&&&&&&&&&&&&&&&&&&&&&
\subsection{Impact of LHC results on Dark Matter searches}
\label{sec:DM}
%&&&&&&&&&&&&&&&&&&&&&&&&&&&&&&&&&

\begin{figure}[htbp]
\begin{center}
\includegraphics[height=7cm]{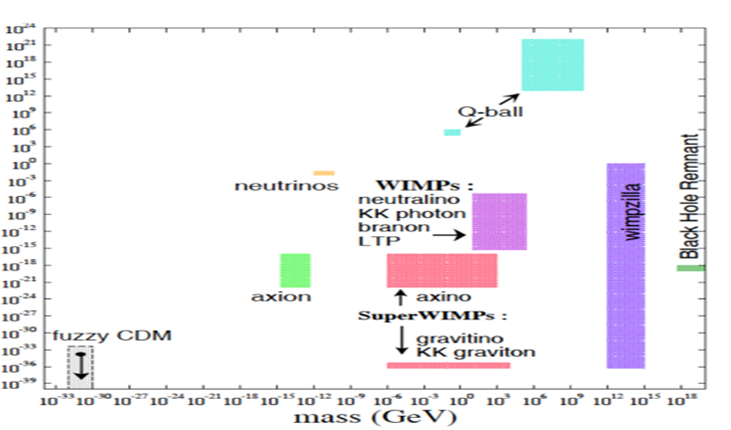}
\end{center}
\caption{DM candidate particles shown in the plane $\chi$-nucleon cross-section (pb) versus $\chi$ mass~\protect\cite{DMCandidatesPlot}.}
\label{fig:DMGene1}
\end{figure}

Dark matter (DM) is required to form the observed large scale structures of the universe and is one of most serious challenge for the Standard Model of Particle Physics. 
The five requirements for a particle to be a dark matter candidate $\chi$
can be spell out as: gravitationally interacting at cosmological and astrophysical scales (the only actual proof that DM exists), not short lived, not hot, 
not baryonic and giving at most the right thermal relic density as measured by CMB experiments, $\Omega h^2=0.120\pm0.003$, where $h$ is the Hubble constant. The 
second condition removes all SM particles apart from the neutrinos and the fermions ($e,u,d$). The third condition rejects the neutrinos and the fourth one 
the rest of the fermions. So SM does not provide any viable DM candidate while ``natural" BSM theories does, 
as shown in Fig.~\ref{fig:DMGene1}. Assuming that dark matter is explained by only one particle with mass $m_{\mathrm{DM}}$ and the 
relevant gauge coupling constant $g_{\mathrm{DM}}$, then $\Omega_{\chi} \propto m_{\mathrm{DM}}^2/g_{\mathrm{DM}}^4$. With this in mind, 
three categories can be formed: $i)$ the WIMP sector where 
$m_{\mathrm{DM}} \approx \Lambda_{\mathrm{EW}}$ and $g_{\mathrm{DM}}=g_{\mathrm{EW}}=2(\sqrt{2}G_F)^{1/2}m_W \approx 0.65$, $ii)$ the hidden sector, gathering 
SuperWIMP and axions~\footnote{The name of the hypothetical particle resolving the strong CP problem.}, where 
$m_{\mathrm{DM}} \leq \Lambda_{\mathrm{EW}}$ and $g_{\mathrm{DM}} \ll g_{\mathrm{EW}}$ and $iii)$ the undetectable sector with fuzzy dark 
matter, where the interaction is purely gravitational. Among all candidates, WIMP particles are still the most popular since they are motivated by the resolution 
of the hierarchy problem, a completely uncorrelated reason (the so-called ``WIMP miracle").

\begin{figure}[htbp]
\begin{center}
\includegraphics[height=7cm]{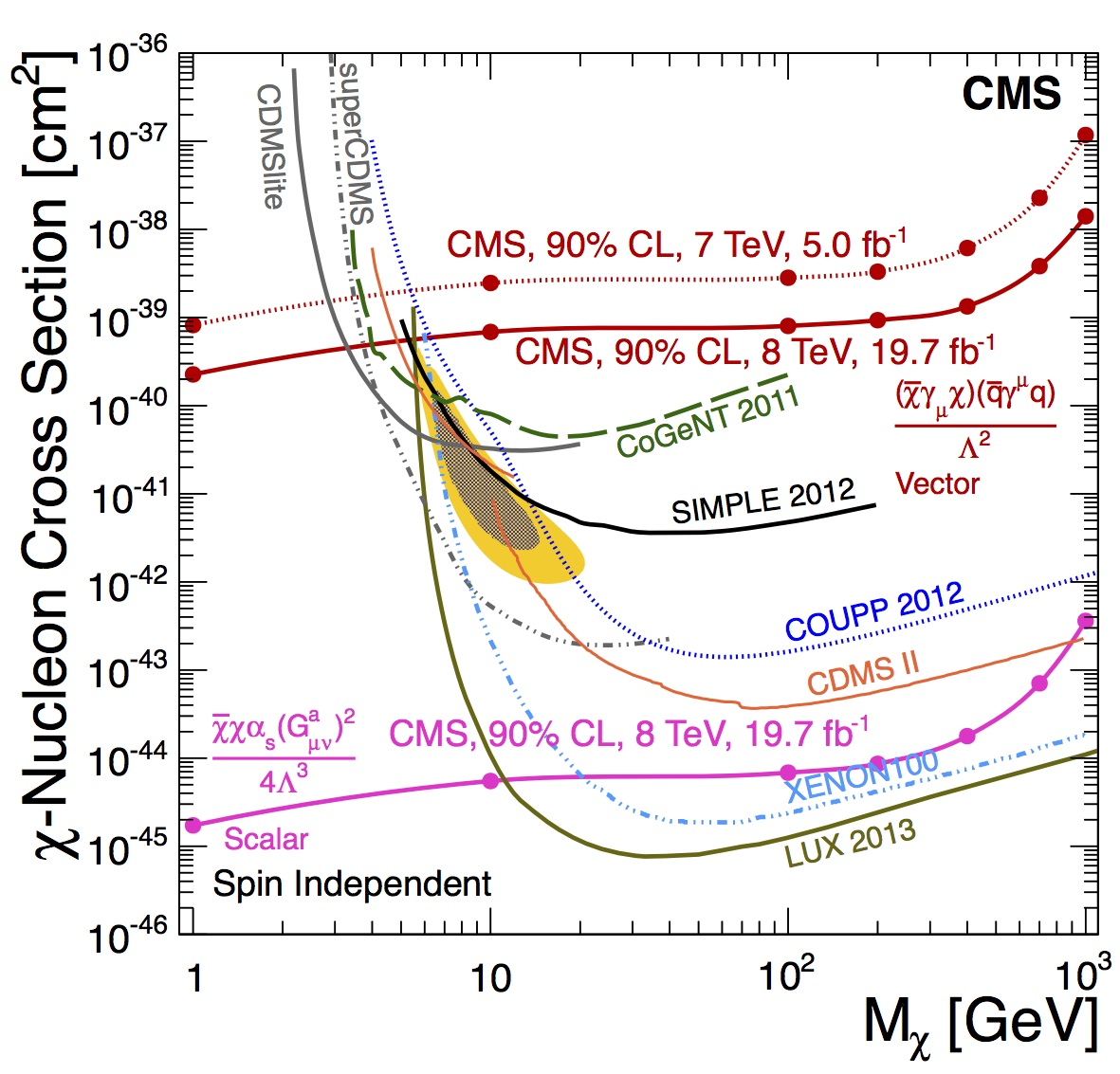}
\end{center}
\caption{Exclusion curves obtained in the same plane by LHC and direct DM search experiments, assuming a vector-like mediator for the $\chi$-nucleon interaction~\protect\cite{CMS_MonoJet}.}
\label{fig:DMGene2}
\end{figure}

Since a lot of information is already available in other lectures of this school~\cite{JSilk} and in excellent 
reviews~\cite{DMCandidates1,DMCandidates2}, I'll only discuss the LHC input to the DM search. By analogy with the weak interaction 
described by Fermi theory, DM could be produced at LHC via $q\bar{q},qg, gg \rightarrow X \rightarrow \chi \bar{\chi}$ and could be observed in a monojet analysis (the 
jet is an initial state radiation) see Fig.~\ref{fig:ED_ADD1}. The mediator $X$ of mass $M$ could be scalar, vector or axial-vector, 
and interact with quark and WIMP with coupling factors $g_{q,g}$ and $g_{\chi}$. The contact interaction scale is then defined as 
$\Lambda=M/\sqrt{g_{q,g} g_{\chi}}$. This approach allows the conversion into DM-nucleon cross-section limits for a given $\chi$ mass, directly comparable with dedicated 
DM searches~\cite{DM_CollLimit1,DM_CollLimit2}. For vector-like mediator, Fig.~\ref{fig:DMGene2} shows that LHC could exclude low mass WIMP where 
direct searches have no sensitivity because of undetectable energy recoil of the nucleon. For axial-vector mediator or DM-gluon coupling with scalar or vector 
mediators, LHC results exceed all present limits~\cite{CMS_MonoJet}.

\begin{figure}[htbp]
\begin{center}
\includegraphics[height=7cm]{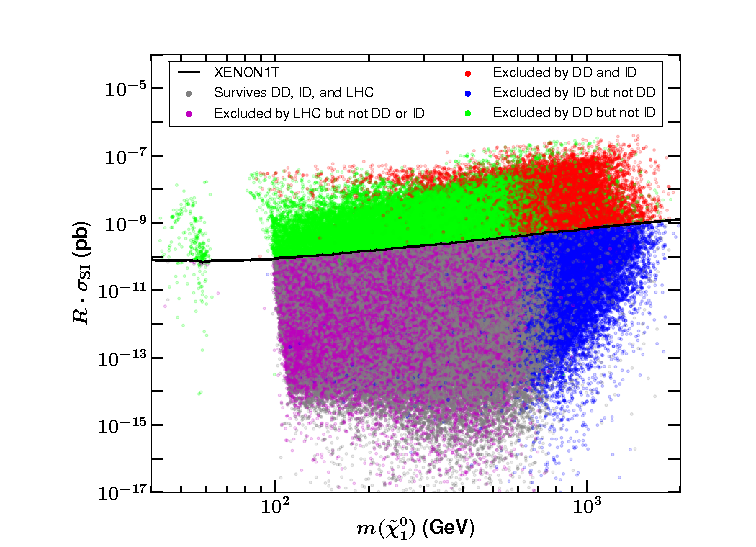}
\end{center}
\caption{MSSM models and experimental constraints in the $\chi$-nucleon cross-section-$m_{\tilde{\chi}_1^0}$ plane~\protect\cite{pMSSM_DM}.}
\label{fig:DMGene_pMSSM1}
\end{figure}

\begin{figure}[htbp]
\begin{center}
\includegraphics[height=7cm]{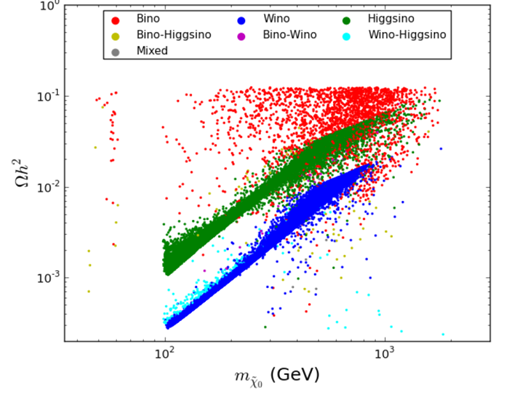}
\end{center}
\caption{LSP composition of surviving pMSSM models in the thermal relic density-$m_{\tilde{\chi}_1^0}$ plane~\protect\cite{pMSSM_DM}.}
\label{fig:DMGene_pMSSM2}
\end{figure}

The leading WIMP candidate is (still) $\tilde{\chi}_1^0$, but it can not be probed by monojet analysis because of the too low cross-section, 
$\sigma=\mathrm{O}(1)~\mathrm{fb}$ for $m_{\tilde{\chi}_1^0}=100~\mathrm{GeV}$. In direct SUSY searches $m_{\tilde{\chi}_1^0}$ is generally not 
directly accessible, but many models predicting $m_{\tilde{\chi}_1^0} \simeq$ 500-600~GeV are currently excluded (Section~\ref{sec:BSM_SUSY}). 
It is therefore interesting to scan a subset of MSSM models, satisfying the present experimental LHC constraints from direct SUSY searches and Higgs mass, 
and see what flavor and mass range of $\tilde{\chi}_1^0$ survives~\cite{pMSSM_DM}. Constraints from
direct and indirect DM non-LHC searches can also be included, as mentioned above and the complementarity of the different approaches appears clearly 
in Fig.~\ref{fig:DMGene_pMSSM1}. The surviving models are shown in Fig.~\ref{fig:DMGene_pMSSM2}. Interestingly 
a huge quantity of models is still alive today and the only models which saturate the thermal relic density have bino-like LSP. 
Note also that almost all surviving models will be reachable by experiments in a near future. To conclude, it is worth to 
mention that the $ZH(\rightarrow \chi \chi)$ searches, not included in this study, are also excellent probes for $m_{\mathrm{DM}}<m_H/2$, and covers  
$\sigma_{\chi-\mathrm{Nucl}}\sim$10$^{-7}$-10$^{-11}$ pb depending on the mediator properties~\cite{Higgs_DM}.

%&&&&&&&&&&&&&&&&&&&&&&&&&&&&&&&&&
\subsection{Neutrinos and Baryogenesis}
\label{sec:neutr}
%&&&&&&&&&&&&&&&&&&&&&&&&&&&&&&&&&

Because neutrinos only interact with matter by exchanging very massive $W$ and $Z$, LHC experiments can not explore the neutrino sector. 
Instead dedicated experiments are built near nuclear plants or in deep underground mines where atmospheric, solar or intense neutrino beams from a particle physics 
center could be studied. Last 15 years saw many new results: discovery of neutrino oscillation~\cite{NeutrinoOsc}, neutrino-tau~\cite{NTAUDISC} and 
recently measurement of third PMNS mixing angle $\theta_{13}$~\cite{NeutrinoTheta}.

Before going further, let's recall the peculiar position of the neutrinos in the SM: $i)$ the only neutral fermions, $ii)$ the only particles with unknown masses, 
present upper limits from direct measurement gives $m<2$ eV~\cite{NeutrinoMass}, $iii)$ the only fermions with no right-handed partners, $iv)$ the only sector 
where original SM setting, $m_{\nu} =0$ and no mixing, was incorrect, $v)$ the only fermions giving a cosmic background (C$\nu$B), expected 
at T=1.95 K$\sim$0.17 meV. On top of this singular situation in the SM four fundamental questions remain unanswered:
Are neutrinos Majorana or Dirac fermions? What is the absolute mass scale of neutrinos and what is their hierarchy (normal or inverted)? 
What are the precise values of PMNS matrix elements, and especially the CP violation phase? And finally are there ``sterile" neutrinos, i.e. neutrino interacting 
only with the Higgs and other lepton doublets but not $W$ or $Z$?

While waiting for experimental answers, the question of the very low neutrino masses compared to other fermions 
triggered very interesting theoretical developments. The preferred explanation relies currently on the so-called ``see-saw" mechanism pioneered at the end of the 
70's~\cite{SeeSaw1,SeeSaw2,SeeSaw3,SeeSaw4}. The plain vanilla scenario assumes Majorana neutrinos ($\bar{\nu}=\nu$) and predict 3 new particles, the 
right-handed sterile neutrino singlets $N_{Ri}$. Since $N_{R}$s do not interact with gauge fields, very high 
masses $M\gg\Lambda_{\mathrm{EW}}$ are possible. Meanwhile left-handed neutrinos need to be massless to conserve the gauge invariance. With left and 
right-handed neutrinos, the Higgs field generates neutrino Dirac masses $m_{D,i}=v h_i$ as for quarks and other leptons. The neutrino mass matrix can therefore be 
written $\bigl(\begin{smallmatrix} 0&m_D\\ m_D&M \end{smallmatrix} \bigr)$ with two neutrino eigenvalues $m_{\nu}=m_D^2/M$ and $m_{N}=M$. 
Assuming very low left-handed neutrino masses $m_{\nu}\sim0.1$ eV and $h_i\sim 1$ gives very heavy right-handed neutrinos $m_N=10^{14-15}$ GeV.
Under these assumptions, a very interesting byproduct can be derived. It has been known since the end of the 70's that the presence of GUT-scale mass particles 
could be a natural way to generate matter-anti matter asymmetry~\cite{GUT_Baryogenesis}. Later, it was realized that the very high mass right-handed 
neutrinos could play this role, assuming the reheat temperature is higher than $N_R$ masses~\cite{Leptogenesis}, see~\cite{VanillaLeptogenesis} for a recent review. 
Indeed in this case the three Sakharov conditions~\cite{Sakharov} are satisfied:
\begin{itemize}
\item For $T<m_N$, $N_R$ will be out of equilibrium (Sakharov 3)
\item $N_R \rightarrow l^+ H$ and $N_R \rightarrow l^- H$ decays violate the leptonic number conservation. CP violation in the lepton sector is 
expected from the PMNS matrix as a single complex phase like for the quark sector in the CKM matrix (Sakharov 2).
\item The lepton asymmetry can be converted to baryon asymmetry by non perturbative SM processes called sphaleron (Sakharov 1).  
\end{itemize}
This scenario is particularly 
popular since: $i)$ the EW baryogenesis is not possible in the SM with the recently measured Higgs mass (Section~\ref{sec:Higgs_Universe}), 
$ii)$ CP violation in the quark sector is not large enough to generate the observed baryon asymmetry~\cite{CKM_Baryogenesis} and $iii)$ it 
turns natural if weak-scale SUSY is realized since $N_R$ will have a superpartner with similar mass. Note also 
that many other scenarios exist as the see-saw mechanism is viable down to very low values of $h_i$. For example for 
$h_i$=$10^{-6,-8}$, the lightest subGeV $N_R$ will be long-lived, a good dark-matter candidate and could explain baryogenesis~\protect\cite{Shaposhnikov}.

%&&&&&&&&&&&&&&&&&&&&&&&&&&&&&&&&&
\subsection{The future of experimental Particle Physics}
\label{sec:prosp}
%&&&&&&&&&&&&&&&&&&&&&&&&&&&&&&&&&

Neutrino and collider experiments have paved the way of particle physics in the last 50 years. Prospects in the next decades are now briefly 
discussed, focussing on new colliders addressing the energy frontier. A more thorough review can be found in~\cite{ESG}. 
Collider experiments must address the two central questions of particle physics: detail understanding of the recently discovered scalar, i.e. 
precise measurement of all Higgs couplings, and thorough search for BSM particles in the TeV scale range. The next step is obviously the 
LHC restart at $\sqrt{s}=13$-14 TeV in 2015. By 2018 (2022), 50 (300) fb$^{-1}$ of data should be collected. New particles with masses augmented 
by a factor 1.5-2 compared to present limits should be accessible. By 2022, a 5-15\% precision on all Higgs couplings, except $c$-quark Yukawa coupling,  
should be at hand~\cite{MPeskin}. Beside this point, three projects are in competition:
\begin{itemize}
\item A High Luminosity LHC (HL-LHC) in 2024-2030 whose aim is to obtain 3000 fb$^{-1}$ of data at $\sqrt{s}=14$ TeV. Here also substantial 
improvements could be obtained in the Higgs sector as well as a first measurement of Higgs self-coupling.
\item A linear e$^+$-e$^-$ collider (ILC) that could start in Japan before 2030 with a $\sqrt{s}=250$ GeV and serve as a Higgs Factory. Full program 
includes an increase at 500 and 1000 GeV. Extra improvements on Higgs coupling precision by factors 2 to 10, depending on the particle type, could be achieved.
\item A circular e$^+$-e$^-$ collider (LEP3) with $\sqrt{s}=240$ GeV could also be a very powerful Higgs Factory with similar or better 
sensitivity. The idea is to dismount the LHC, install LEP-like set-up in place, and run from 2024 on.
\end{itemize}
Beyond 2030, higher energy machines (VLHC, CLIC and TLEP) could be the continuation of the three previous projects with $\sqrt{s}=$26-100, 0.5-3.0, 
and 0.24-0.35 TeV respectively. These programs require new tunnels in the CERN area: 80 km for VLHC and TLEP and 13-48 km for CLIC.

\begin{figure*}[htbp]
\begin{center}
\includegraphics[height=6cm]{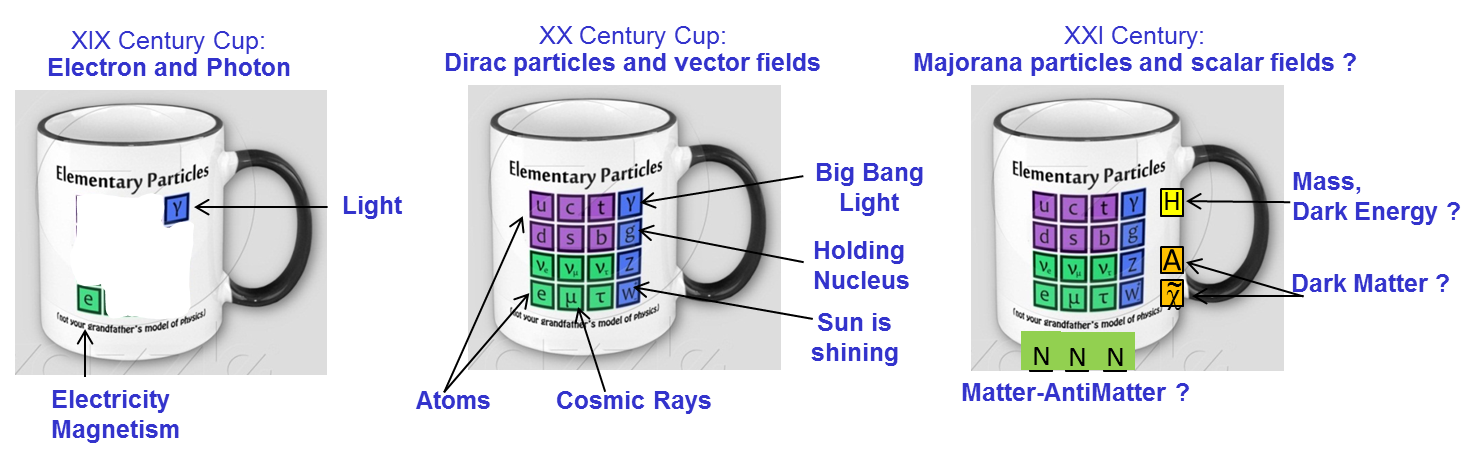}
\end{center}
\caption{Past, present and a possible future of particle physics.}
\label{fig:Conclu}
\end{figure*}

Aside of these projects, new types of colliders are being developed, based on electron plasma activated by a laser. The main advantage of 
this completely new technology for particle physics is to increase the current accelerating gradient by a factor 
1000 with respect to conventional Radio Frequency technology. Reaching 100 GV/m will considerably reduce the collider size~\cite{PlasmaLaser}. If this 
technology continues to follow the Livingstone law (Fig.~\ref{fig:LHC1}) a 1-km TeV collider could be envisaged by 2035 and be  
competitive with CLIC, TLEP and VLHC projects.

%&&&&&&&&&&&&&&&&&&&&&&&&&&&&&&&&&&&&&&&&&&&&&&&&&&&&&&&&&&&&&&&&&&&&&&&&&&&&&&&&&&&&&&&&&&&&&&&&&&&&&&&&&&&&&&&&&&&&&&&&&&&&&&&&&&&&
%&&&&&&&&&&&&&&&&&&&&&&&&&&&&&&&&&
%&&&&&&&&&&&&&&&&&&&&&&&&&&&&&&&&&
%&&&&&&&&&&&&&&&&&&&&&&&&&&&&&&&&&
\section{Conclusions}
\label{Part:Conclusion}
%&&&&&&&&&&&&&&&&&&&&&&&&&&&&&&&&&

Particle physics and cosmology are facing a particularly intriguing moment. Theoretically both are described by Standard Models with few parameters 
(triumph for the principle of simplicity?) and extremely robust against more and more precise experimental 
data. In particle physics, it's been 40 years without BSM discovery, despite the huge number of models predicting new physics close to the EW scale now 
extensively probed by LHC. In cosmology, $\Lambda$CDM is still a good fit despite the reduction of allowed parameter space volume by 
$10^5$ during the last 15 years. Experimental findings are even more tantalizing: the cosmological constant is very small but 
not 0 (1998), the SM Higgs seems to exist at a mass of 125 GeV and is apparently fine-tuned 
(2013). Both are pointing away from naturalness, even if the latter is still fresh and needs the full LHC program to be really conclusive.

Despite this quite unique situation for physics, the future should be paved by the understanding of current puzzles, i.e the nature of dark energy, 
dark matter and matter-antimatter asymmetry. Figure~\ref{fig:Conclu} left/center shows that the discovery of new particles has always been a way to answer 
fundamental questions. So may be the XXI$^{\mathrm{rst}}$ century will continue the tradition as suggested by the right mug ... and ultimately 
decide whether our universe is natural or not ?

%&&&&&&&&&&&&&&&&&&&&&&&&&&&&&&&&&
\vspace*{0.5cm}
\textbf{Acknowledgments}
%&&&&&&&&&&&&&&&&&&&&&&&&&&&&&&&&&
I'd like to express my gratitude to the organizers of the 100$^{\mathrm{th}}$ Les Houches Summer School for giving me the opportunity to 
deliver this lecture. It was also a real pleasure for me to stay some days in Les Houches and profit from the stimulating audience 
of cosmologists in a beautiful landscape.

\bibliographystyle{atlasnote}
\bibliography{LesHouchesPralavorioArXiv}

\end{document}